\RequirePackage{fix-cm}
\documentclass[onecolumn,final]{svjour3}

\def\makeheadbox{\relax}

\usepackage[authoryear]{natbib}
\usepackage[utf8]{inputenc}
\usepackage{textcomp}
\usepackage{textgreek}
\usepackage{amsmath,bm}
\usepackage{amssymb}
\usepackage{graphicx}
\usepackage{mathabx}  
\usepackage{float}
\usepackage{color}
\usepackage[caption=false]{subfig}
\usepackage{soul} 

\usepackage{nomencl}
\usepackage{ifthen}
\renewcommand\nomgroup[1]{%
  \ifthenelse{\equal{#1}{A}}{%
    \item[\textbf{Acronyms}]}{
  \ifthenelse{\equal{#1}{D}}{%
    \item[\textbf{Dimensionless Quantities}]}{
  \ifthenelse{\equal{#1}{R}}{%
    \item[\textbf{Roman Symbols}]}{
  \ifthenelse{\equal{#1}{G}}{%
    \item[\textbf{Greek Symbols}]}{
  \ifthenelse{\equal{#1}{S}}{%
    \item[\textbf{Superscripts}]}{
  \ifthenelse{\equal{#1}{U}}{%
    \item[\textbf{Subscripts}]}{
  \ifthenelse{\equal{#1}{X}}{%
    \item[\textbf{Other Symbols}]}{
  {}}}}}}}}}

\makenomenclature

\definecolor{gray1}{gray}{0.6}

\journalname{Experiments in Fluids}
\begin{document}

\title{Refractive index matching (RIM) using double-binary liquid-liquid mixtures}

\author{Thorben Helmers  \and Philip Kemper \and Ulrich Mießner \and Jorg Thöming}

\institute{Thorben Helmers
\at Department of Environmental Process Engineering, University Bremen, Leobener Str. 6, 28359 Bremen/Germany\\Tel.: +49-421-63337, Fax: +421-218-98 63337\\\email{helmers@uvt.uni-bremen.de}
\and Philip Kemper
\at Department of Chemical Process Engineering, University Bremen, Leobener Str. 6, 28359 Bremen/Germany\\Tel.: +49-421-63466, Fax: +421-218-98 63466\\\email{kemper@uni-bremen.de}
\and Ulrich Mießner 
\at Department of Environmental Process Engineering, University Bremen, Leobener Str. 6, 28359 Bremen/Germany\\Tel.: +49-421-63333, Fax: +421-218-98 63333\\\email{miessner@uvt.uni-bremen.de}
\and Jorg Th\"oming
\at Department of Chemical Process Engineering, University Bremen, Leobener Str. 6, 28359 Bremen/Germany\\Tel.: +49-421-63300, Fax +49-421-63302\\\email{thoeming@uni-bremen.de}
}

\date{2019}

\maketitle
\begin{abstract}
Within the last decade microscopic multiphase flows have gained broad interest. An exact understanding of the underlying hydrodynamic interrelations is the key for successful reactor layout and reaction control. To examine the local hydrodynamic behavior, non-invasive optical measurements techniques like particle tracking velocimetry (PTV) or (micro)particle image velocimetry ((\textmu)PIV) are the method of choice, since they provide precise velocity measurement with excellent spatial resolution. Such optical approaches require refractive index matching (RIM) of the involved flow phases to prevent optical distortion due to light refraction and reflection at the interfaces. Established RIM approaches often provide a single one degree of freedom which is sufficient to match the RI of the flow phases solely. With that, the material properties ($Oh$ number) are fixed and the relevant dimensionless numbers ($Ca$, $Re$) may only be altered hydrodynamically or geometrically. To avoid expansive geometric scaling of the microchannels, we propose an approach using two binary mixtures (double-binary mixtures) to introduce an additional degree of freedom. The approach allows examining liquid-liquid two-phase flows at a distinct velocity while being able to change the material combination ($Oh$-Number). Therefore $Ca$ and $Re$ can be chosen individually and the RIM provides undisturbed optical access. We present 4 different binary mixtures to be used, e.g. with Taylor droplets. The relevant material parameters are successfully correlated to measurement data, which delivers a system of equations that determines the mass fractions and the velocities to address $Re$ and $Ca$ individually. A proof-of-principle for the proposed double binary mixture RIM-approach is successfully established using \textmu PIV raw images.      

\end{abstract}

\renewcommand{\nomname}{List of Symbols} 

\printnomenclature[5em]

\section{Introduction and Concept}\label{sec:intro}

Microfludic processes using liquid-liquid multiphase flows has gained great interest in the last decades \citep{Zhao.2011,Chou.2015, Ahmed.2018,Shi.2019}. The applications range from chemical \citep{Kobayashi.2006,Song.2006,Kralj.2007,Lang.2012,Tanimu.2017} to biological \citep{ClausellTormos.2008,Mazutis.2013,Chen.2014,Wolf.2015,Hosokawa.2017} and pharmaceutical applications \citep{Kang.2014,Piao.2015}. The large specific surface area in combination with well defined and allegedly easy-to-control flow structures promise a great potential for process intensification. Thus, optimal reactor design and controllability of the flow is key to raise this potential.  

Deep insight into the hydrodynamic interaction of the distinct flow phases enables to tune e.g. the mass transfer rates towards more sustainable operation modes close to the optimal working point \citep{Magnaudet.2000,Ern.2012}. To achieve this, numerical simulation and CFD-calculations enable relatively easily a parameterized study of e.g. liquid-liquid Taylor flows \citep{LuisA.M.Rocha.2017}. Dimensionless numbers ($Re$, $Ca$) and material property ratios ($Oh$) can be set freely and independently to identify e.g. critical operation modes. 

The dimensionless quantities are based on the material parameters of the continuous phase since it provides the wall contact and drives the flow of the disperse phase. The capillary number compares the viscous forces with the interfacial tension forces $Ca = \frac{u_0 \eta_c}{\sigma}$ and it is based on the superficial velocity $u_0$, the dynamic viscosity $\eta_c$ of the continuous phase and the interfacial tension $\sigma$. The Reynolds number relates the inertia forces to the viscous forces $Re = \frac{\rho_c u_0 d}{\eta_c}$, where $\rho_c$ denotes the density of the continuous phase and $d$ the diameter of the microchannel. The Ohnesorge number $Oh = \sqrt{\frac{Ca}{Re}}$ removes the hydrodynamic influences and solely remains as a material parameter. Please note that for macroscopic liquid-liquid flows, the set of dimensionless variables changes with the growing influence of inertia and gravitational forces to $Re$, $We$ and $Mo = Bo\,Oh^2$. Where the Weber number $We = Ca\,Re$ represents the inertia forces compared to the interfacial tension forces. $Mo$ introduces the buoyancy dependence of the flow's material system \citep{Araujo.2012} when combining the Bond number $Bo = \frac{\Delta\rho\,g\,d^2}{\sigma}$ with the Ohnesorge number. Since buoyancy forces are small ($Bo \ll 1$) in microfluidic applications, we base our RIM system on the $Oh$ number only.

Preferably non-invasive experimental methods need to be applied to investigate the hydrodynamic behavior of liquid-liquid multiphase flows supporting the numerical findings. Even in microchannels, a high spatial resolution of an entire flow field is accessible with e.g. optical measurement techniques \citep{Park.2006,Kinoshita.2006, Khodaparast.2013}. However, undistorted optical access is necessary to avoid measurement deviation due to light refraction and reflection. 

Often the hardware related refraction effects of the setup can be compensated for adjusting the experimental design e.g. avoiding curved surfaces and using of corrective optics. The curved interfaces of microscopic liquid-liquid flows are commonly counteracted applying refractive index matching with one degree of freedom \citep{Miessner.2008,Ma.2014,Liu.2017}. A broad overview of possible liquid-liquid, as well as solid-liquid refractive index matching possibilities, is given in the works of \citet{Budwig.1994} and \citet{Wright.2017}. Recently, several approaches have been made using refractive index matched systems to mimic special application cases like specific rheology for blood \citep{Najjari.2016,Brindise.2018}, high-density differences or for a buoyant jet \citep{Clement.2018, Krohn.2018}.

In microscopic liquid-liquid flows RIM with one degree of freedom allows matching the RI of one phase solely to the other. The material system of the phases like density, viscosity and interfacial tension are fixed for the desired RI. The governing dimensionless numbers such as $Re$, $Ca$ may only be parameterized hydrodynamically (superficial flow velocity) or geometrically (microchannel diameter). The monetary effort to parameterize the diameter of the microchannel is high, while the velocity alters both quantities simultaneously. 

Alternatively, surfactants could be added to change the interfacial tension of the material system solely. Surfactant concentration well below the critical micelle concentration does not significantly change the viscosity of the host phase. Albeit this approach indeed addresses $Ca$ only, the use of surfactants introduces severe effects such as altering the hydrodynamics of the flow as well as its mass transfer properties. 

We suggest an approach using two immiscible binary mixtures to match refractive indices \citep{Saksena.2015,Cadillon.2016}, i.e. adding an additional degree of freedom to the system by introducing a binary mixture for each of the liquid phases. The now flexible material system provides an entire range of RI to match the immiscible binary liquid mixtures optically. Hence, the simultaneous velocity related change of $Ca$ and $Re$ may be compensated for by adequately adapting the material composition of the mixtures ($Oh$).

We carefully determine the mass fraction dependent material properties of the involved mixtures, establish dedicated correlation functions and provide an optimization algorithm to calculate the necessary information to use this RIM system. A proof of principle is given using \textmu PIV raw-images of two fluorescence particle-seeded microscopic Taylor flows at two $Re$ and constant $Ca$ at different channel heights.

Recently, the viscosity ratio of both flow phases $\lambda$ is reported to influence the local hydrodynamics of microscopic Taylor droplets \citep{Rao.2018}. However, their measurements are limited to a narrow parameter range without refractive index matching \citep{Kovalev.2018, Liu.2017}. Therefore, we suggest four combinations of well quantified double-binary mixtures for the disperse as well as the continuous phase. This allows to additionally alter the viscosity ratio of the flow phases by changing the flow system. 
 
\nomenclature[R]{$u$}{velocity [$mm \cdot s^{-1}$]}
\nomenclature[U]{$tot$}{total}
\nomenclature[U]{$0$}{superficial}
\nomenclature[G]{$\eta$}{dynamic viscosity [$Pa \cdot s$]}
\nomenclature[G]{$\sigma$}{interfacial tension [$N \cdot m^{-1}$]}
\nomenclature[D]{$Ca$}{Capillary number [$-$]}
\nomenclature[U]{$c$}{continuous phase}
\nomenclature[U]{$d$}{disperse phase}
\nomenclature[U]{$n$}{nonpolar phase}
\nomenclature[U]{$p$}{polar phase}
\nomenclature[A]{\textmu PTV}{particle tracking velocimetry}
\nomenclature[A]{\textmu PIV}{particle image velocimetry}
\nomenclature[A]{DMSO}{dimethyl sulfoxide}
\nomenclature[A]{RIM}{refractive index matching}
\nomenclature[D]{$Re$}{Reynolds number [$-$]}
\nomenclature[G]{$\rho$}{density [$kg \cdot m^{-3}$]}
\nomenclature[R]{$d$}{characteristic length [$m$]}
\nomenclature[D]{$\lambda$}{Viscosity ratio [$-$]}
\nomenclature[D]{$Oh$}{Ohnesorge number [$-$]}
\nomenclature[D]{$Bo$}{Bond number [$-$]}
\nomenclature[D]{$Mo$}{Morton number [$-$]}
\nomenclature[R]{$g$}{gravitational acceleration [$m \cdot s^{-2}$]}
\nomenclature[G]{$\xi$}{mass fraction [$kg \cdot kg^{-1}$]}
\nomenclature[R]{$A$}{correlation factor [$-$]}
\nomenclature[U]{$1,2,3$}{order of correlation factor }
\nomenclature[U]{$ex$}{excitation [$-$]}
\nomenclature[G]{$\lambda$}{light wavelength [$nm$]}
\nomenclature[A]{$NA$}{numerical aperture}
\nomenclature[A]{$DOF$}{depth of field}

\section{Material and Methods}\label{sec:doub-bin:methods}
Within this section, the choice of the basic mixture compounds, the experimental procedures to retrieve the properties of the fluids and the numerical approach are described. All measurements are referenced to mass fractions $\xi$ to compensate non-linearities (e.g., excess volume while mixing or when preparing solutions).

\subsection{Basic Mixture Compounds}\label{sec:doub-bin:basicmix}

For the binary mixtures of this work, we focused to mainly use nontoxic, non-hazardous, newtonian substances. A wide range of addressable refractive indices is accomplished, when mixing substances if high and and low RI in each phase. The polar phase is chosen to be aqueous for practical reasons. The range of RI of the aqueous binary mixture is defined when using either DMSO or glycerol as newtonian liquid to elevate the RI. The nonpolar phase is based on hexane, as it represents a tradeoff between a lower RI and hazardous properties. The nonpolar binary mixture is complemented by either anisole or sunflower-oil, which establishes a high RI range. Rheometry measurements show, that in the observed range sunflower-oil also behaves newtonian.

Combining the suggested aqueous mixtures properties with the non-polar systems, the viscosity ratio $\lambda$ between the phases of the flow can be changed by either inverting the flow phases or changing the combination of the materials. For that reason, we introduce and characterize different binary mixtures for the polar as well as for the non-polar phase. Please note, the viscosity ratio cannot be chosen freely. However, it is possible to choose a ratio below or above unity to investigate the hydrodynamic consequences. 

For low viscous mixtures with viscosity ratio below one, we use water/DMSO and n-hexane/anisole, while for high viscosity ratio we propose to apply water/glycerol and n-hexane/sunflower oil.  

With these mixtures, we reach a broad range of refractive indices, laying between the pure substance's RI of n-hexane ($RI = 1.3753$) and glycerol/DMSO ($RI \approx 1.47$). The material properties of the binary mixtures (density, viscosity, interfacial tension), as well as the refractive, are given in Sec. \ref{sec:doub-bin:results}.

\subsection{Determination of material properties}\label{sec:doub-bin:meas}
The \textbf{refractive indices} of the used fluids are measured using a Kr\"uss Abbe refractometer AR2008, temperated at 20\textdegree C. The refractive indices were measured using a light wavelength of 589\,nm (Sodium D1-line) with three independent measurements each. The refractometer uses a liquid film between two prisms to measure the angle of total reflection, from which the refractive index is calculated. For volatile or liquids with low viscosity or low surface tension, no stable film forms between the prisms, which leads to a high deviation. For this reason, the refractive indices of the n-hexane/anisole system for high mass fractions of anisole could not be acquired reliably. Therefore literature data is used.

The \textbf{interfacial tension} of the double binary mixtures is measured using a Lauda TVT drop tensiometer with a syringe of 2.5\,ml and a stainless steel capillary at a temperature of 20\textdegree C. The interfacial tensions are acquired using the volume drop method: A droplet is formed at a capillary tip. Based on the balance of buoyancy and surface tension the droplet pinches off into the continuous phase. The dispersed droplet is detected with a light barrier and the interfacial tension calculated from the droplet volume.

The interfacial tension is a time-dependent property for nonpure or mixed systems. Thus, as part of the measurement of the dynamic interfacial tension, droplets with differently aged interfacial area are formed. The interfacial tension for infinite time $\sigma_{\inf}$ can be derived depending only on bulk diffusion of surface-active substances. Therefore the interfacial tensions are correlated following \citet{Wilkinson.1972} and \citet{Sinzato.2017}. Within our work, the droplet formation times and therefore the surface age lie between 2.5\,s and 60\,s.

The light barrier, that detects the detachment of the ascending droplets, cannot work properly when the refractive indices are matched. Thus, we deliberately detune the mass fraction of the nonpolar phase such, that a sufficient detection difference between the refractive indices of both phases ($\Delta RI=0.01$) exists. The droplet detachment, as well as the interfacial tension, are determined at these detuned fluid compositions. The actual interfacial tension for the matched case is then retrieved by linear interpolation using the lever rule. The linear interpolation is considered valid since only the mass fraction of one of both phases is varied and the change in the mass fraction is kept small.

The measurements of the \textbf{viscosities} for the volatile compounds (n-hexane/anisole) are performed using a Malvern Kinexus Ultra Plus with solvent trap, cone-plate setup, 1 degree opening angle and a stationary shear rate table 10\,s\textsuperscript{-1} .. 100\,s\textsuperscript{-1}  with 5\,\% stationarity tolerance. The remaining measurements are performed using a Bohlin Rheometer CS and a 30\,ml double-gap system. All measurements were performed at 20\textdegree C. 

The \textbf{densities} of the mixtures are mostly retrieved from literature data.

\subsection{Calculating mass-fractions and the superficial velocity}\label{sec:doub-bin:optim}
The main merit of the double-binary mixture flow system is the ability to address different $Re$ and $Ca$ independently of each other. Keeping in mind the definition of $Re$ and $Ca$, this can be stated as an optimization problem with the superficial velocity $u_0$ as the target value. The mass fractions of the continuous phase $\xi_c$ and disperse phase $\xi_d$ are the control variable, which influences the material properties. 

If the fluid properties for the pure substances as well as for the mixtures are known, $Ca$ and $Re$ can be calculated. Because the user needs to establish the flow at a distinct $Re$ and $Ca$, the equations can be changed such, that the superficial velocity $u_0$ for both numbers is a function of the continuous phase mass fraction $\xi_c$:

\begin{equation}\label{eq:doub-bin:ZZZ1}
u_{0,Re} = \frac{Re \cdot \eta_{c}\left( \xi_c  \right) }{\rho_{c}\left( \xi_c \right) \cdot d}
\end{equation}

\begin{equation}\label{eq:doub-bin:ZZZ2}
u_{0,Ca} = \frac{Ca \cdot \omega\left(  \xi_c  \right)}{\eta_{c}\left( \xi_c \right)}
\end{equation}

\nomenclature[U]{$s$}{slug}

The problem defines as a minimization problem to find $u_0$, where the difference of $u_{0,Ca} - u_{0,Re} \approx 0$. Depending on the material property function (given in Sec. \ref{sec:doub-bin:bin-mix}), the problem is not necessarily strictly monotonous. Therefore the use of numerical solvers is strictly advised. A working example using \textit{fminsearch} from MATLAB is given in the supplementary resources. After input of the desired $Re$ and $Ca$, this program automatically determines the necessary superficial velocity $u_0$ and mass fraction of both phases. For a minimum residuum of this optimization ($u_{0,Re}=u_{0,Ca}$), the correct $u_0$ is known as well as the corresponding mass fraction of the continuous phase $\xi_c$. The mass fraction of the disperse phase $\xi_d$ can be calculated via the refractive index of the continuous phase.

\section{Experimental Results}\label{sec:doub-bin:results}
Within this section, the material properties of the binary and RI-matched double-binary mixtures are experimentally investigated.
\subsection{Properties of binary mixtures}\label{sec:doub-bin:bin-mix}
To solve the optimization problem, a solver needs a steady optimization function to work with. However, the measurement data consists of discrete points including a measurement error: A transition of the discrete data sampling to continuous functions is necessary. Thus, we correlate the data with polynomial approaches for simplicity.

The equation for a polynomial function follows

\begin{equation}\label{eq:doub-bin:fiteq}
y = A_3 \cdot \xi^3  + A_2 \cdot \xi^2 + A_1 \cdot \xi + A_0
\end{equation}

where $y$ is the desired property (e.g., density, viscosity or interfacial tension), $\xi$ the mass fraction of the mixture and $A_3$-$A_0$ fitting coefficients for the polynomial. The polynomial coefficients for the different material properties as well as the experimental data and the range of validity are shown in the following. 

\begin{figure*}[htb] 
	\centering
	\subfloat[][]{\includegraphics[width=0.4\linewidth]{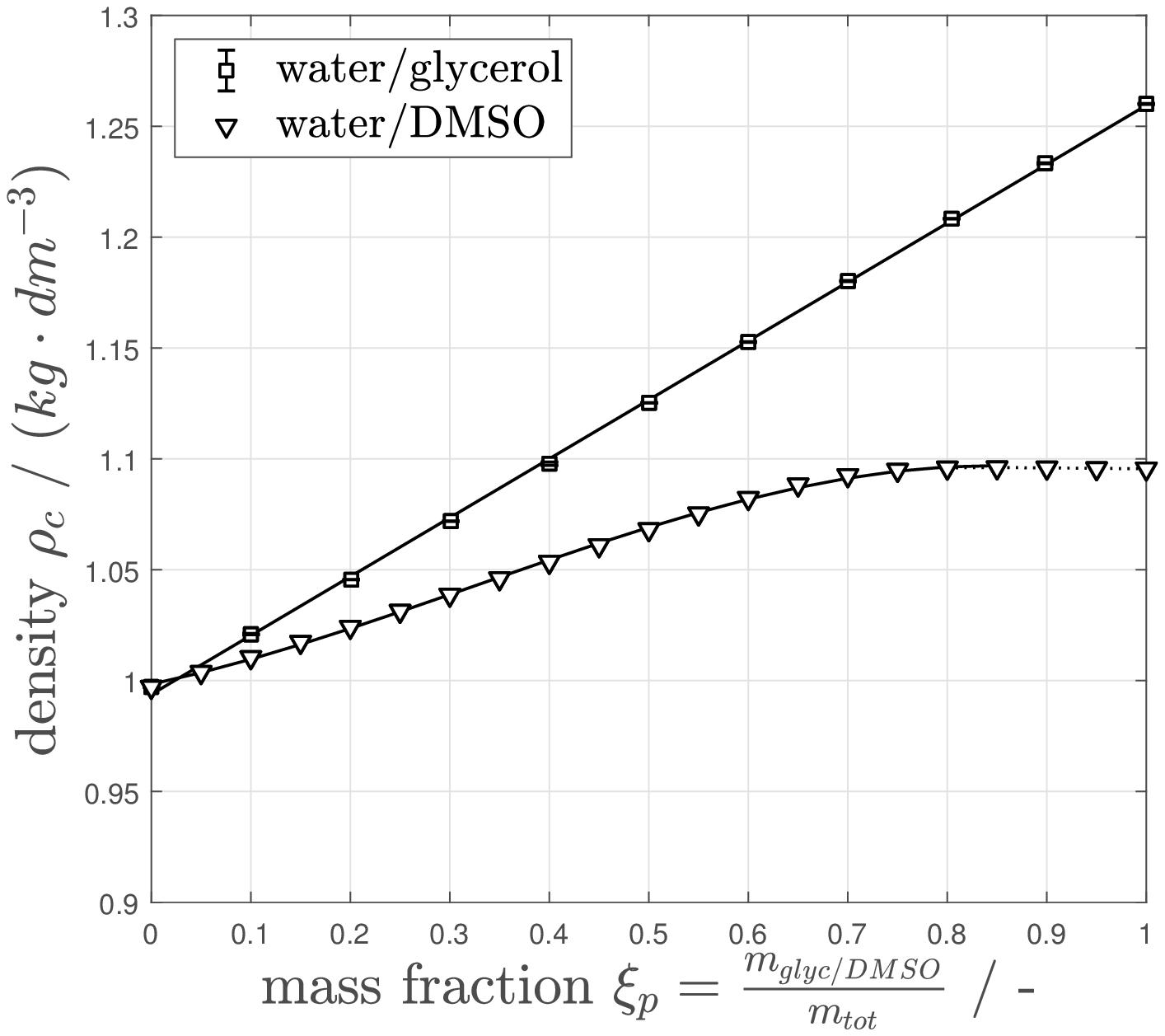}}%
	\qquad
	\subfloat[][]{\includegraphics[width=0.4\linewidth]{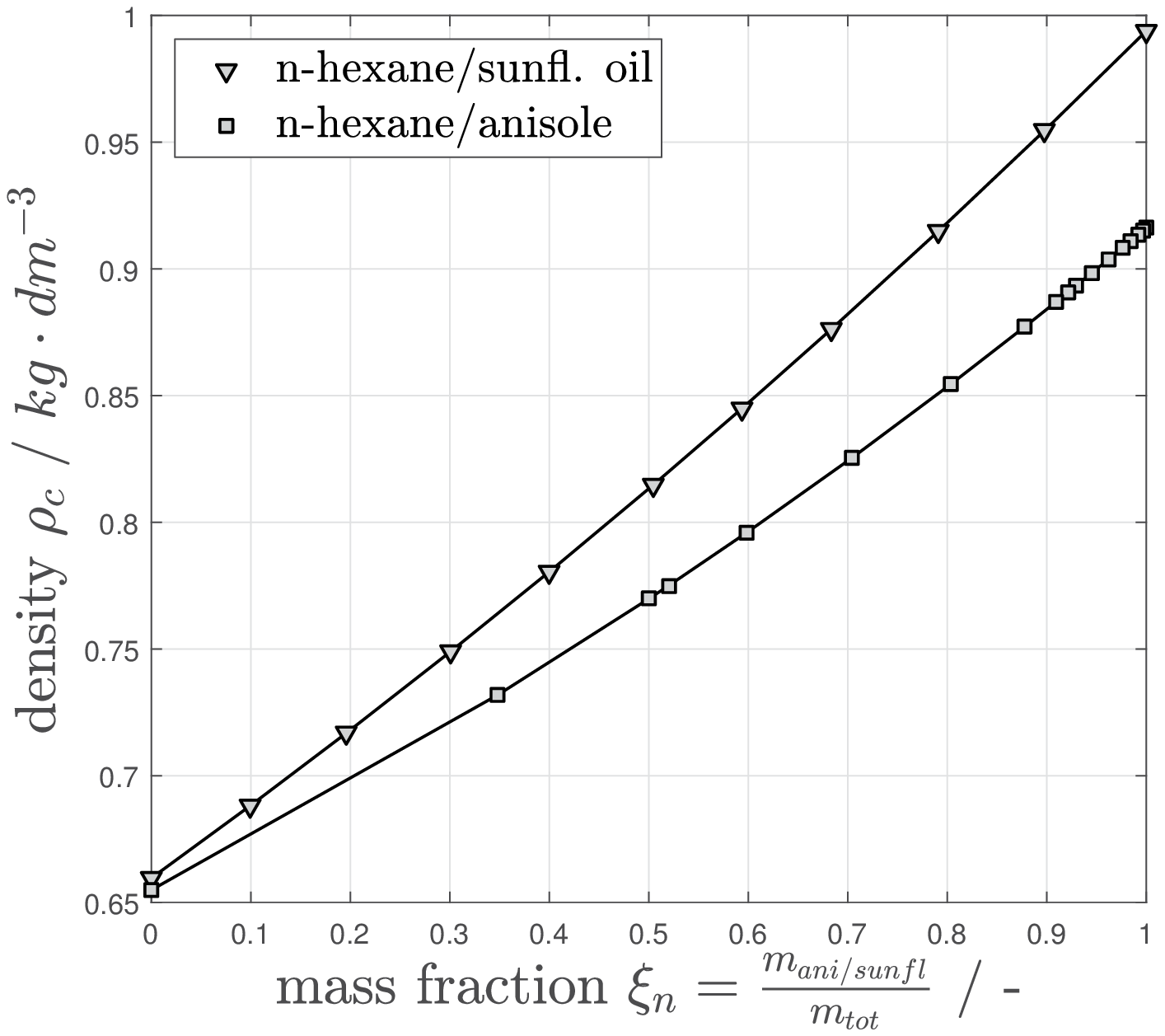}}%
	\caption{Densities for the binary fluid mixtures of each phase a) polar phase for the system water/glycerol (empty squares, own measurements) and water/DMSO (empty downside triangles, \citet{LeBel.1962})
		b) non-polar phase for the system n-hexane / anisole (filled downside triangles, \citet{AlJimaz.2005} and n-hexane/sunflower oil (filled squares, \citet{Gonzalez.1996})}
	\label{fig:doub-bin:dens}
\end{figure*}

At first, the \textbf{densities} of the binary mixtures are investigated in Fig. \ref{fig:doub-bin:dens} a) for the polar and in Fig. \ref{fig:doub-bin:dens} b) for the nonpolar media. The system water/glycerol shows a nearly linear behavior, while the system water/DMSO shows a peak at approximately $\xi = 0.8$. This is caused by nonlinear mixing behavior due to the similarity of water/DMSO and also affects the viscosity. The system n-hexane/anisole shows a nearly linear behavior, while n-hexane/sunflower oil shows a deviation from linearity. The correlations extracted from the measurement and literature data are shown in Tab. \ref{tab:doub-bin:corr_dens}.

\begin{table*}[htb]
\centering
	\caption{Correlation coefficients for the densities of binary mixtures}
	\label{tab:doub-bin:corr_dens}       
	\begin{tabular}{l|c|r|r|r|r}
		\hline\noalign{\smallskip}
		\textbf{mixture} & range & $A_3$ & $A_2$ & $A_1$ & $A_0$ \\
		\noalign{\smallskip}\hline\noalign{\smallskip}
		water / glycerol  & $0.00 < \xi_p <1.00$    & 0  & 0 & 0.2657 & 0.9938 \\
		water / DMSO      & $0.00 < \xi_p <0.85$ & -0.1935 & 0.1883 & 0.0958 & 0.9984\\
		& $0.85 < \xi_p <1.00$ & 0.2667 & -0.7371 & 0.6742 & 0.8919 \\
		hexan / anisole   & $0.00 < \xi_n <1.00$    & 0   & 0.129 & 0.2021 & 0.661\\
		hexan / sunflower oil & $0.00 < \xi_n <1.00$ & 0.0081     &0.0507 & 0.2025 & 0.6549\\
		\noalign{\smallskip}\hline
	\end{tabular}
\end{table*}

\begin{figure*}[htb]
	\centering
	\subfloat[][]{\includegraphics[width=0.4\linewidth]{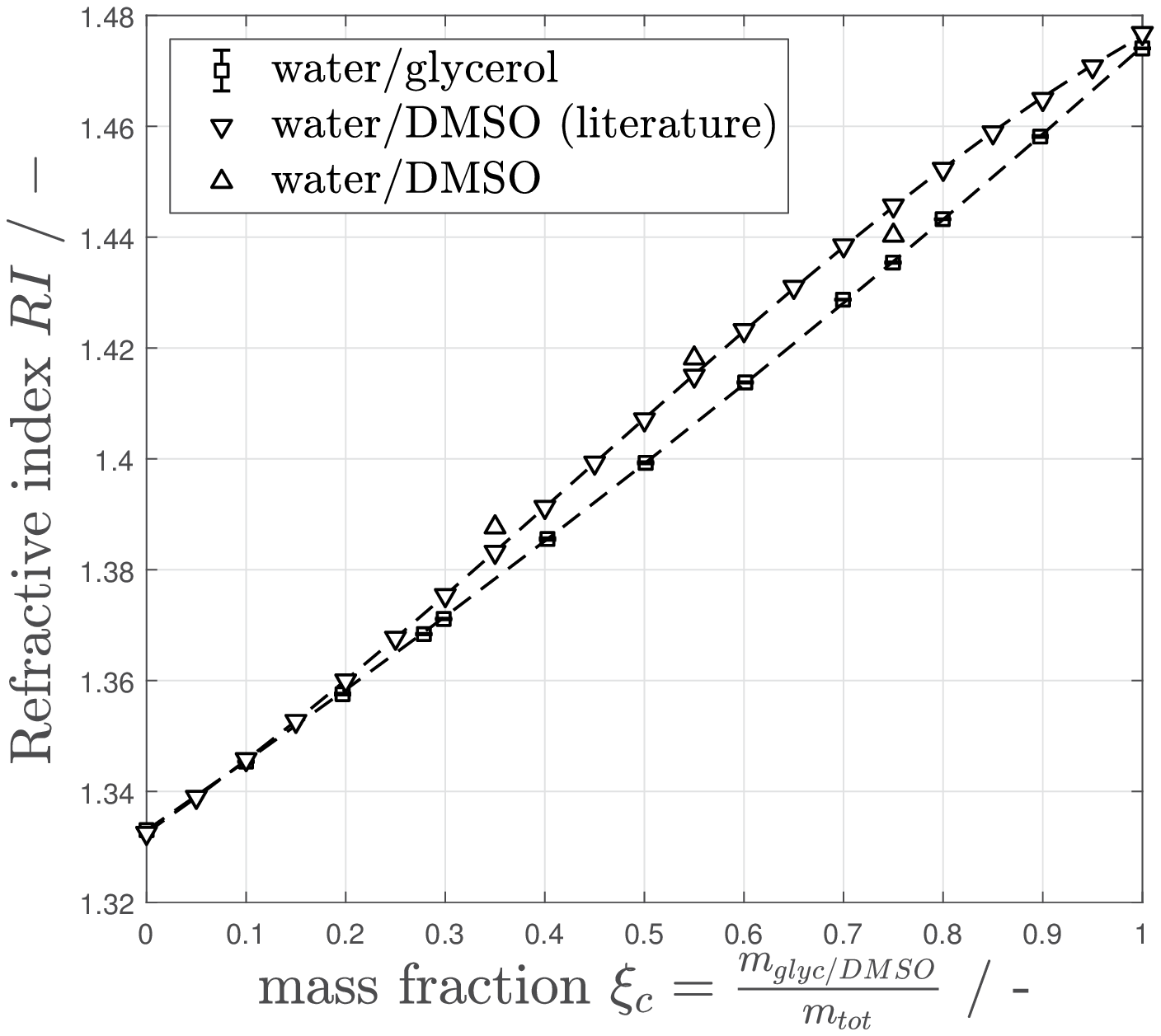}}%
	\qquad
	\subfloat[][]{\includegraphics[width=0.4\linewidth]{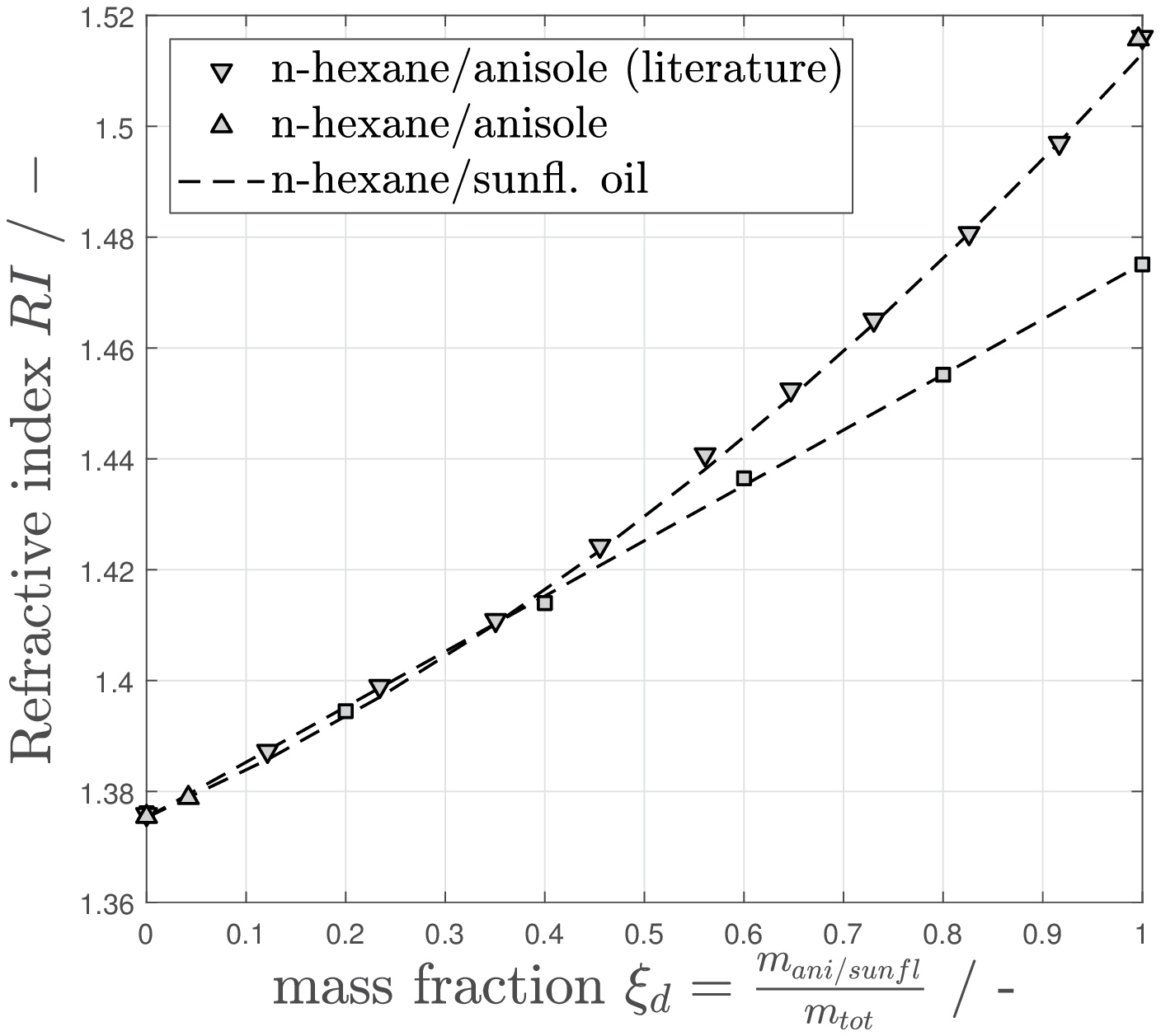}}%
	\caption{Refractive indices for the fluid mixtures a) polar phase for the system water/glycerol (empty squares, own measurements) and water/DMSO (empty downside triangles, \citet{LeBel.1962}, upside triangles own measurements)
		b) non-polar phase for the system n-hexane / anisole (filled downside triangles, \citet{AlJimaz.2005}, upside triangles, own measurements) and n-hexane/sunflower oil (filled squares, own measurements)}
	\label{fig:doub-bin:ri}       
\end{figure*}

In Fig. \ref{fig:doub-bin:ri} the \textbf{refractive index} of the  water/glycerol mixture features a linear behavior (empty squares), while water/DMSO shows a nonlinearity for higher DMSO mass fractions (Fig. \ref{fig:doub-bin:ri} a) , triangles). For the nonpolar substances the system n-hexane/sunflower-oil (filled squares) behaves nearly linear, while n-hexane/anisole (filled triangles) increases the slope indicating non-linear dependence (Fig. \ref{fig:doub-bin:ri} b) ). For all mixtures, the measurements confirm the available literature data. The correlation coefficients are shown in Tab. \ref{tab:doub-bin:corr_RI}.

\begin{table}[h]
	\caption{Correlation coefficients for refractive index}
	\label{tab:doub-bin:corr_RI}       
	\begin{tabular}{l|r|r|r|r}
		\hline\noalign{\smallskip}
		\textbf{mixture} & $A_3$ & $A_2$ & $A_1$ & $A_0$ \\
		\noalign{\smallskip}\hline\noalign{\smallskip}
		water / glycerol  & 0              & 0.0182 & 0.1232 & 1.3300 \\
		water / DMSO       & -0.0631        & 0.0834 & 0.1235 & 1.3300\\ 
		hexan / anisole    & 0              & 0.0588 & 0.0790 & 1.3753\\
		hexan / sunflower oil & 0     &0 & 0.0999 & 1.3753\\
		\noalign{\smallskip}\hline
	\end{tabular}
\end{table}

\begin{figure*}[tb]
	\centering
	\subfloat[][]{\includegraphics[width=0.4\linewidth]{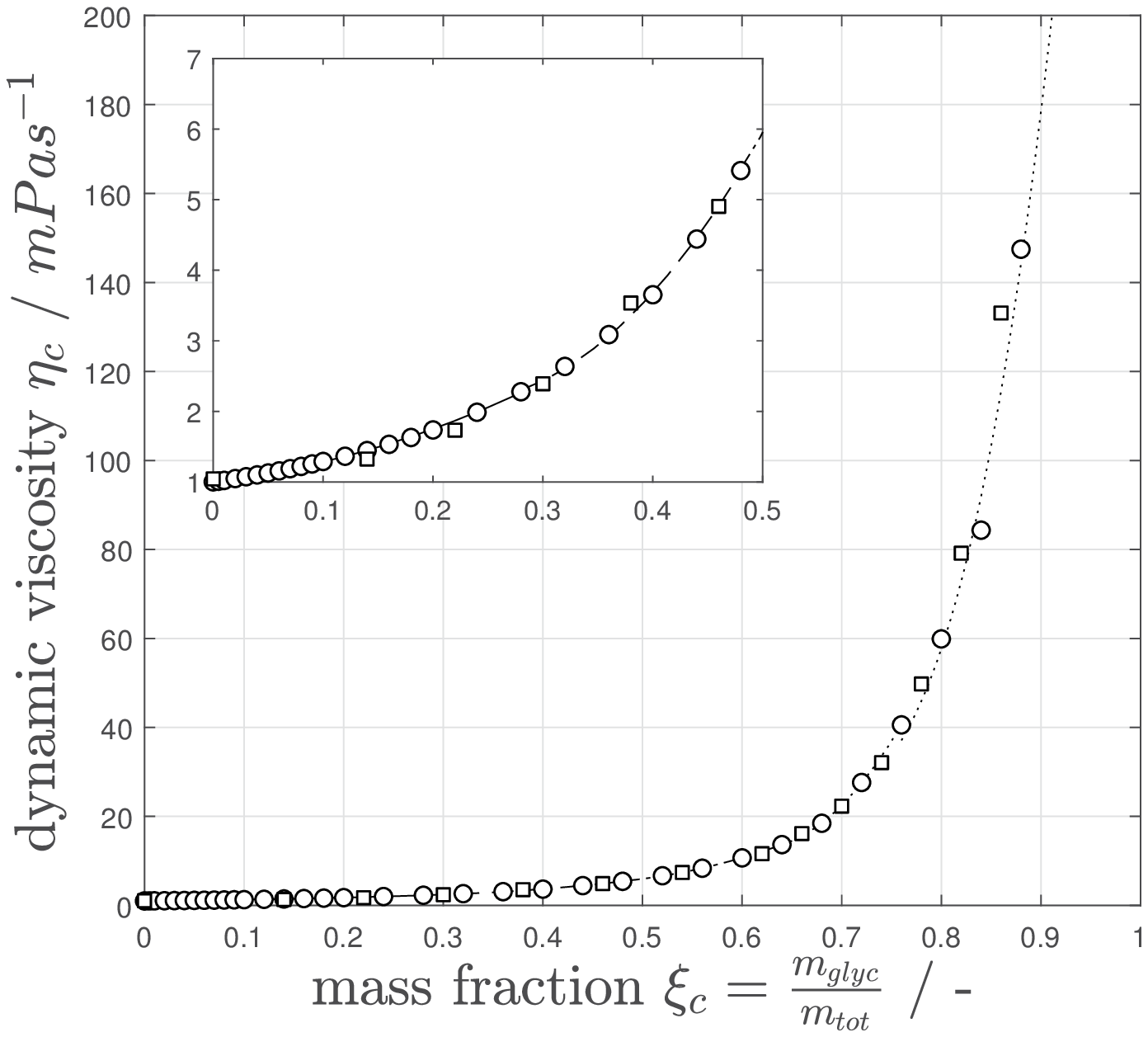}}%
	\qquad
	\subfloat[][]{\includegraphics[width=0.4\linewidth]{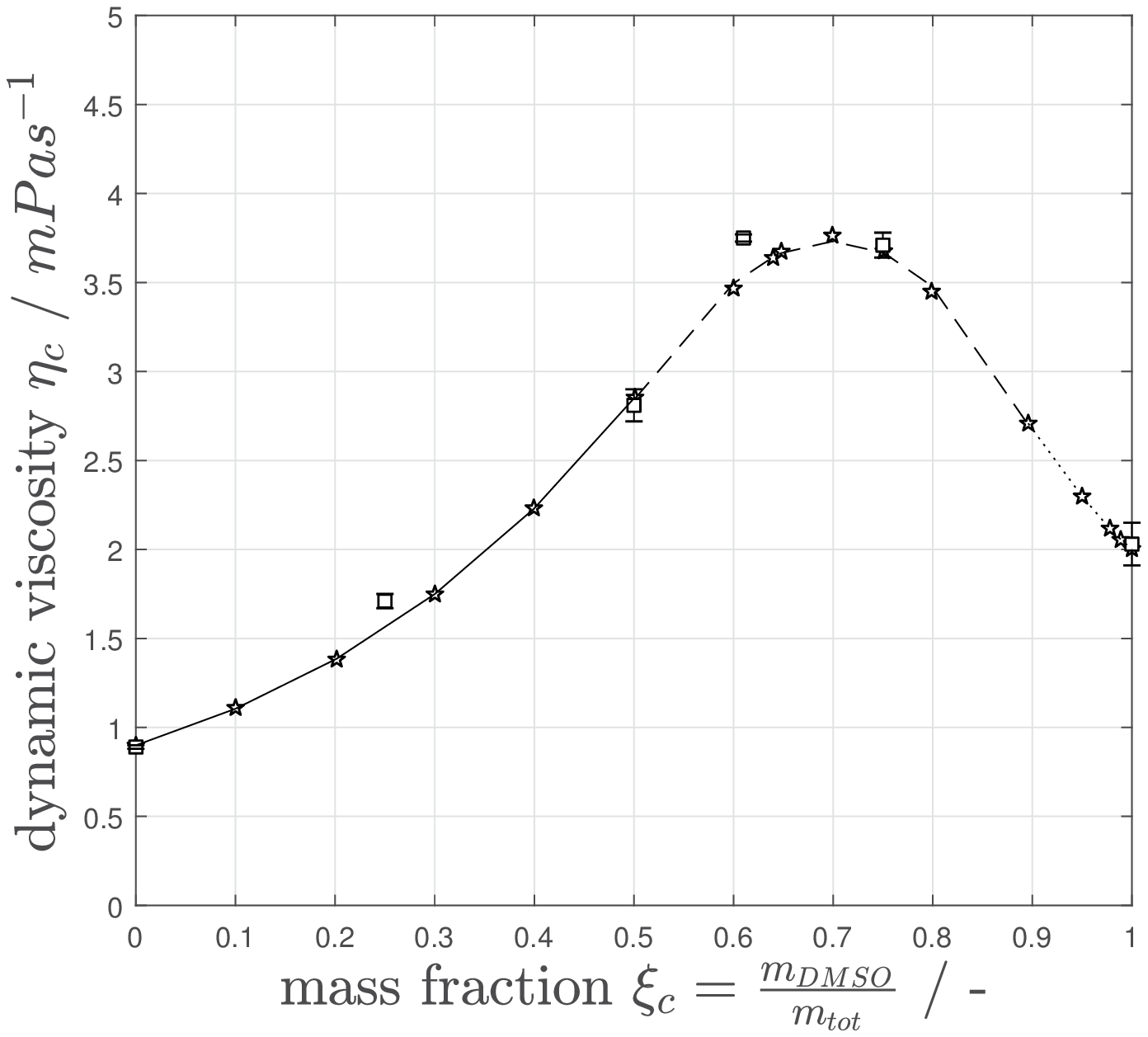}}%
	\qquad
	\subfloat[][]{\includegraphics[width=0.4\linewidth]{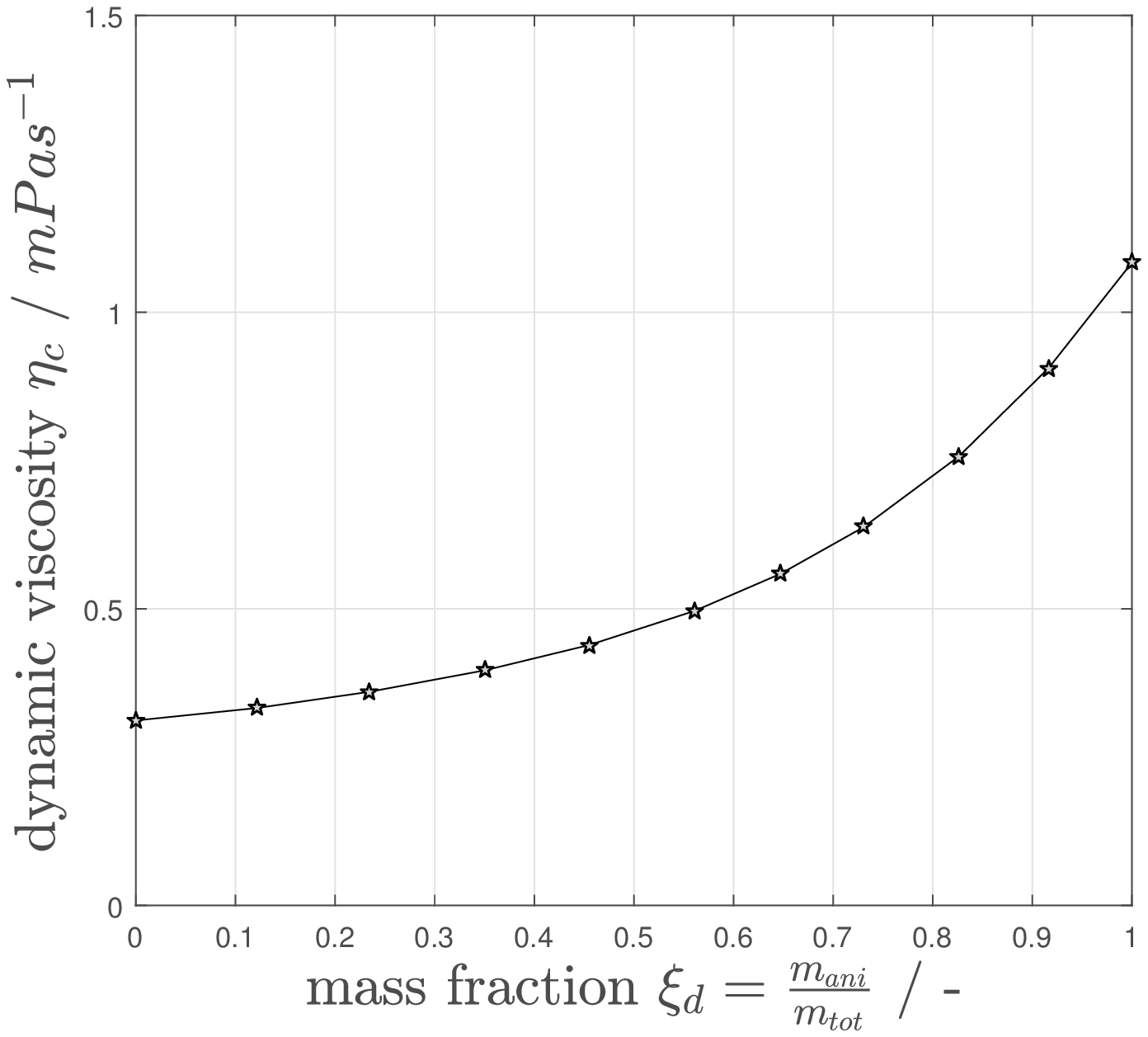}}%
	\qquad
	\subfloat[][]{\includegraphics[width=0.4\linewidth]{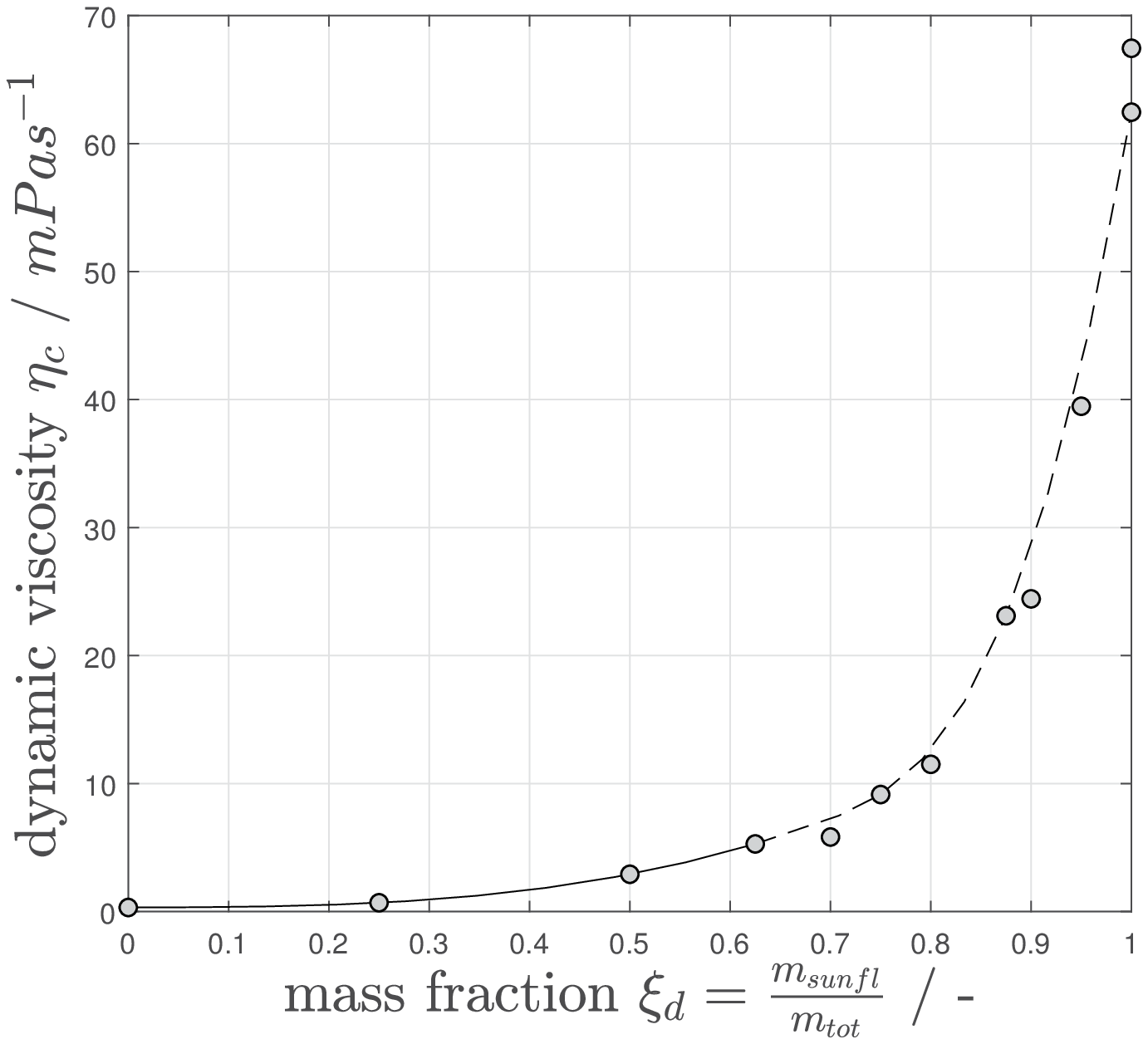}}%
	\caption{Viscosities for the fluid mixtures at 20 \textdegree C a) polar phase water/glycerol (empty circles, \citet{Weast.1989}, empty squares own measurements) b) polar phase water/DMSO (empty stars, \citet{LeBel.1962}, empty squares own measurements)
		c) non-polar phase n-hexane / anisole (filled stars, \citet{AlJimaz.2005})
		d) non-polar phase n-hexane/sunflower oil (filled circles, own measurements)}
	\label{fig:doub-bin:viscos}       
\end{figure*}

The \textbf{viscosities} of the four binary mixtures are given in Fig. \ref{fig:doub-bin:viscos}. Both of the highly viscous mixtures (water/glycerol Fig. \ref{fig:doub-bin:viscos} a) and hexane/sunflower-oil Fig. \ref{fig:doub-bin:viscos} d) ) exhibit similar behavior. As intended, with a higher mass fraction of the more viscous substances, the viscosity rises. 
The less viscous mixtures water/DMSO (Fig. \ref{fig:doub-bin:viscos} b) ) and n-hexane/anisole (Fig. \ref{fig:doub-bin:viscos} c) ) show different behavior. While the viscosity of n-hexane/anisole mixtures increases quadratically with a rising mass fraction of anisole, the viscosity of water/DMSO reaches a peak at $\xi = 0.70$. For n-hexane/anisole no own measurements could be performed, since the high evaporation rate of the volatile hexane/anisole mixture leads to a significant deviation in the mass fraction of the mixture during the measurement. Instead, literature data is used. The correlation coefficients are shown in Tab. \ref{tab:doub-bin:corr_visc}.

\begin{table*}[htb]
	\centering
	\caption{Correlations coefficients for viscosity}
	\label{tab:doub-bin:corr_visc}       
	\begin{tabular}{l|c|r|r|r|r|r}
		\hline\noalign{\smallskip}
		\textbf{mixture} & range & $A_4$ &$A_3$ & $A_2$ & $A_1$ & $A_0$ \\
		\noalign{\smallskip}\hline\noalign{\smallskip}
		water / glyc.  & $0.00 < \xi_p <0.28$     & 0 & 9.6514 & 1.7769 & 1.0095 \\
		& $0.28 < \xi_p <0.48$     & 0 & 0 & 49.223 & -21.944 & 4.5865 \\
		& $0.48 < \xi_p <0.64$     & 0 & 0 & 182.8 & -153.48 & 36.991 \\
		& $0.64 < \xi_p <0.76$     & 0 & 0 & 1272.7 & -1557.1 & 488.84 \\
		& $0.76 < \xi_p <1.00$     & 0 & 10863.4 & -21776.31 & 14652.78 & -3289.78 \\
		water / DMSO      & $0.00 < \xi_p <0.50$     & 0 & 3.6080 & 2.4412 & 1.7748 &  0.8995 \\
		& $0.50 < \xi_p <0.89$    & 0  & -11.9506 & 0.4866 & 16.9060 & -4.2424 \\
		& $0.89 < \xi_p <1.00$     & 0 & 130.6611 & -355.8453 & 315.2710 & -88.0846 \\
		hexan / anisole   & $0.00 < \xi_n <1.00$    & 0.7284  & -0.4838 & 0.3969 & 0.1302 & 0.3117 \\
		hexan / sunfl. oil & $0.00 < \xi_n <0.63$ & 0 & 19.1488 & 0.4864 & 0.1616 & 0.3140 \\
		& $0.63 < \xi_n <1.00$ & 0 & 1309.0133 & -2622.0800 & 1774.9866 & -399.4200 \\
		\noalign{\smallskip}\hline
	\end{tabular}
\end{table*}       

\subsection{Properties of RI-matched double-binary mixtures}
\begin{figure*}[htb]
	\centering
	\subfloat[][]{\includegraphics[width=0.4\linewidth]{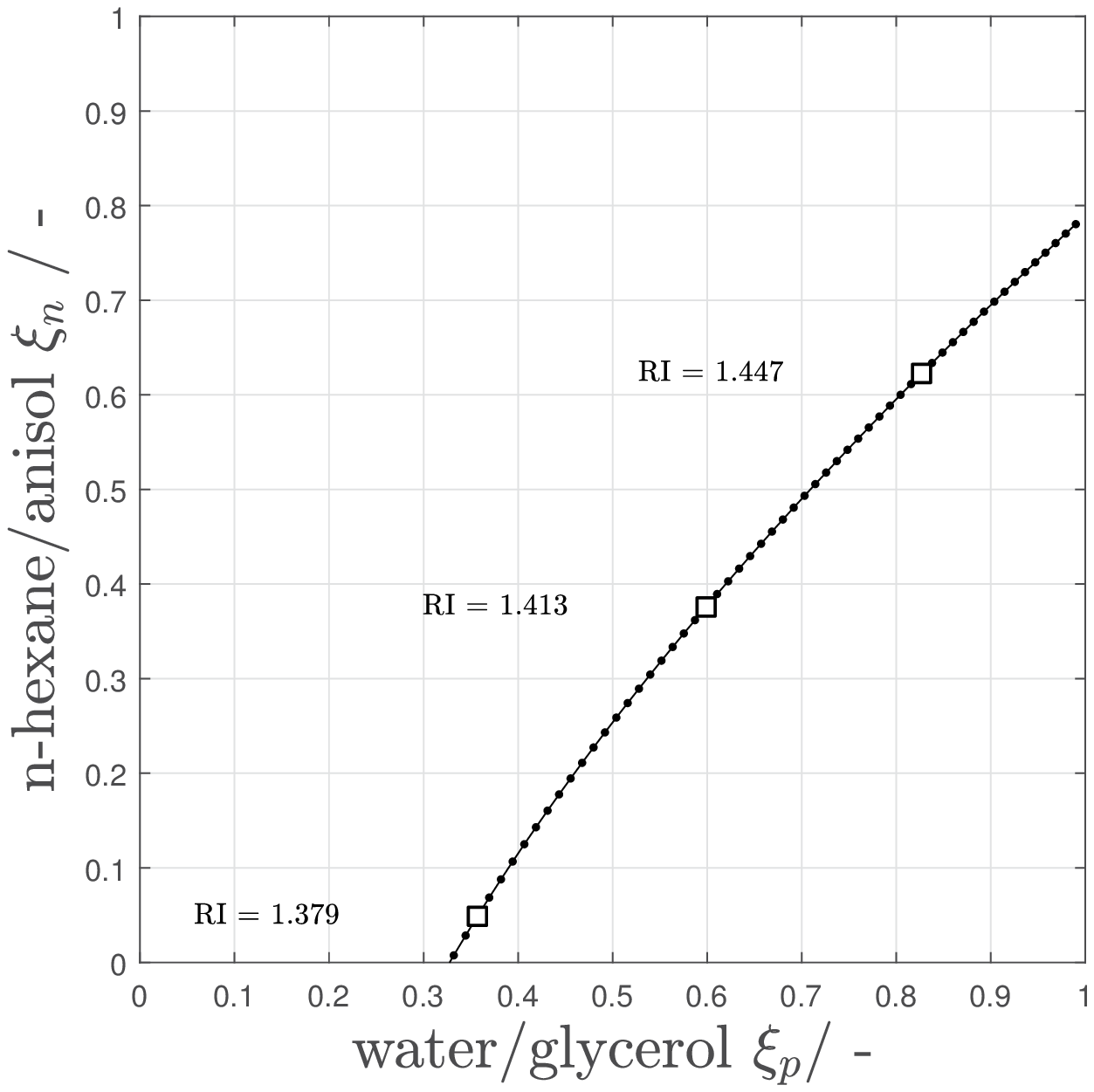}}%
	\qquad
	\subfloat[][]{\includegraphics[width=0.4\linewidth]{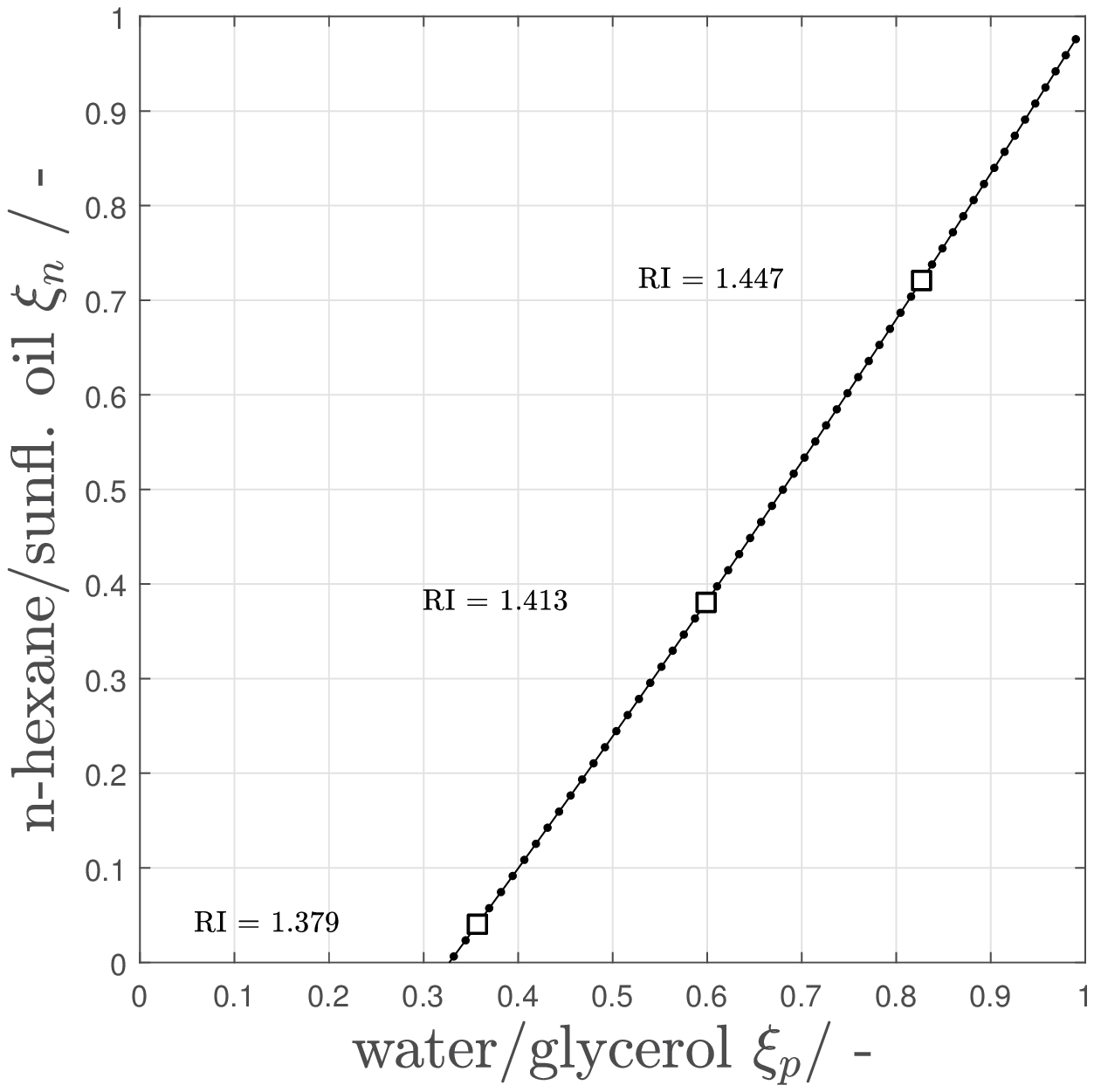}}%
	
	\subfloat[][]{\includegraphics[width=0.4\linewidth]{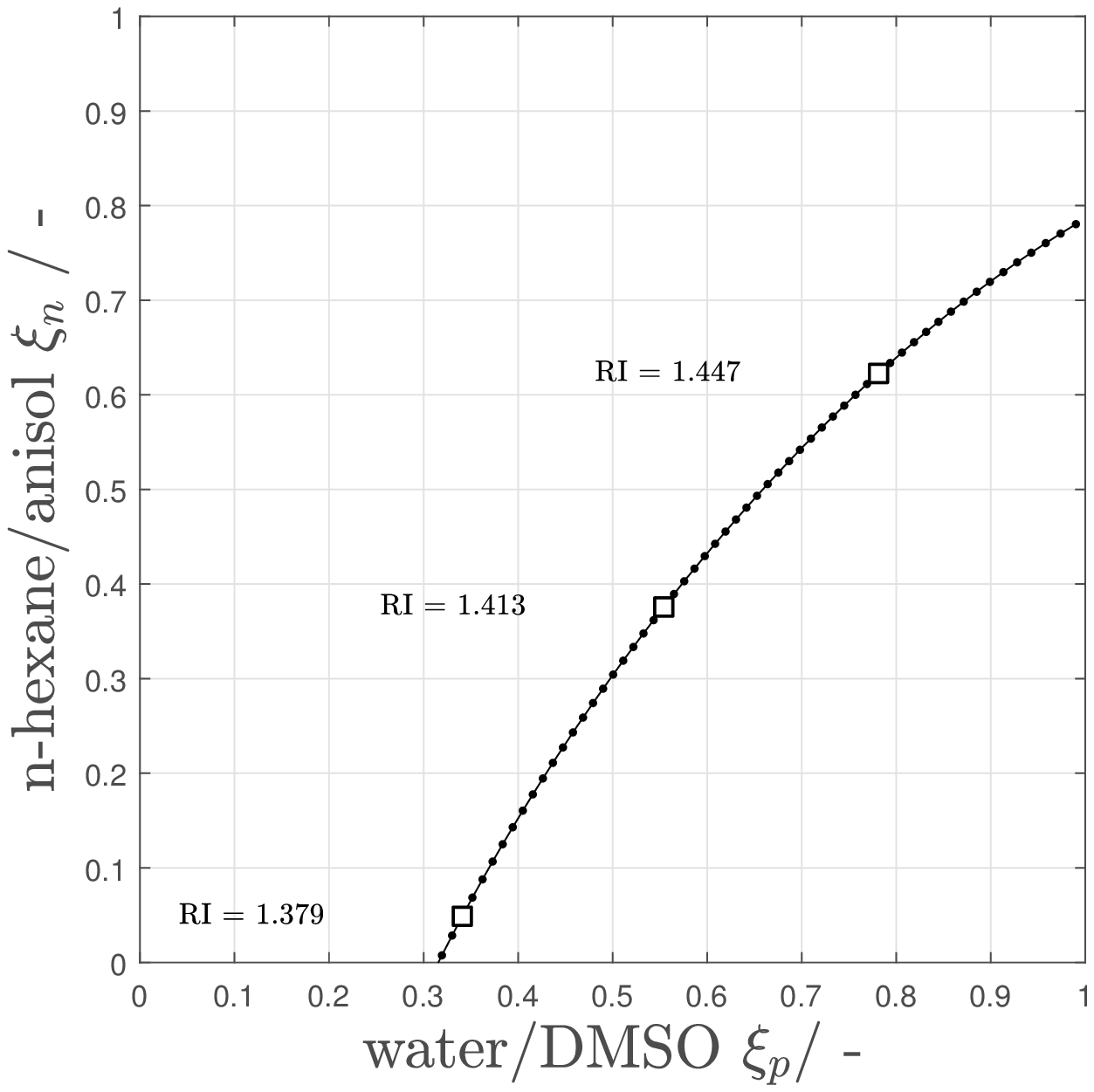}}%
	\qquad
	\subfloat[][]{\includegraphics[width=0.4\linewidth]{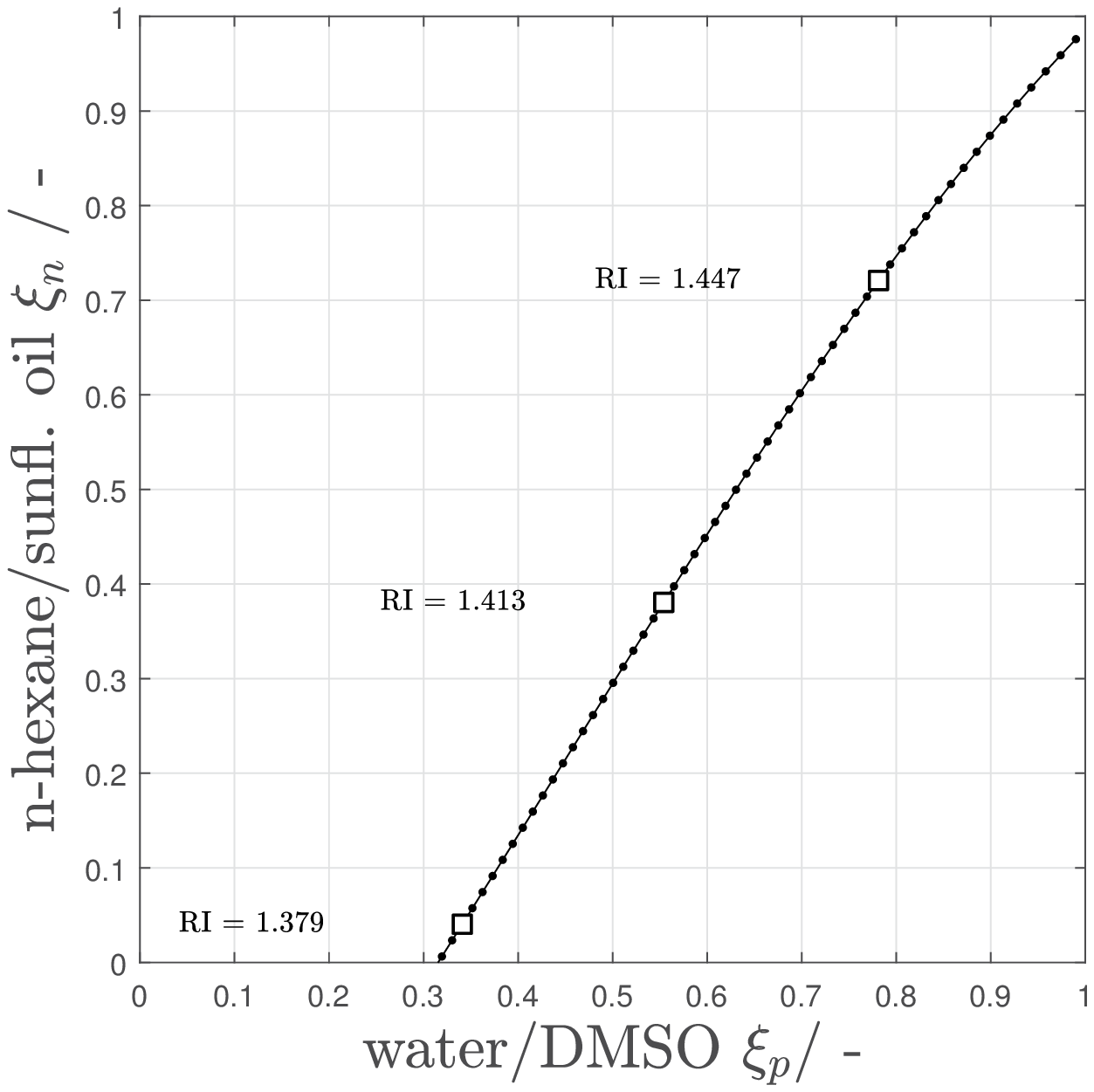}}%
	\caption{Mass fractions of RI-matched double-binary mixtures with matched RI along the graph. a) water/glycerol - n-hexane/anisole b) water/glycerol - n-hexane/sunflower-oil c) water/DMSO - n-hexane/anisole d) water/DMSO - n-hexane/sunflower-oil. Large circles represent a $\Delta RI$ = 0.068 step, small dots the interiorly intermediate steps $\Delta RI$ = 0.0034}
	\label{fig:doub-bin:matched-ri}       
\end{figure*}

\begin{figure*}[h]
	\centering
	\subfloat[][]{\includegraphics[width=0.4\linewidth]{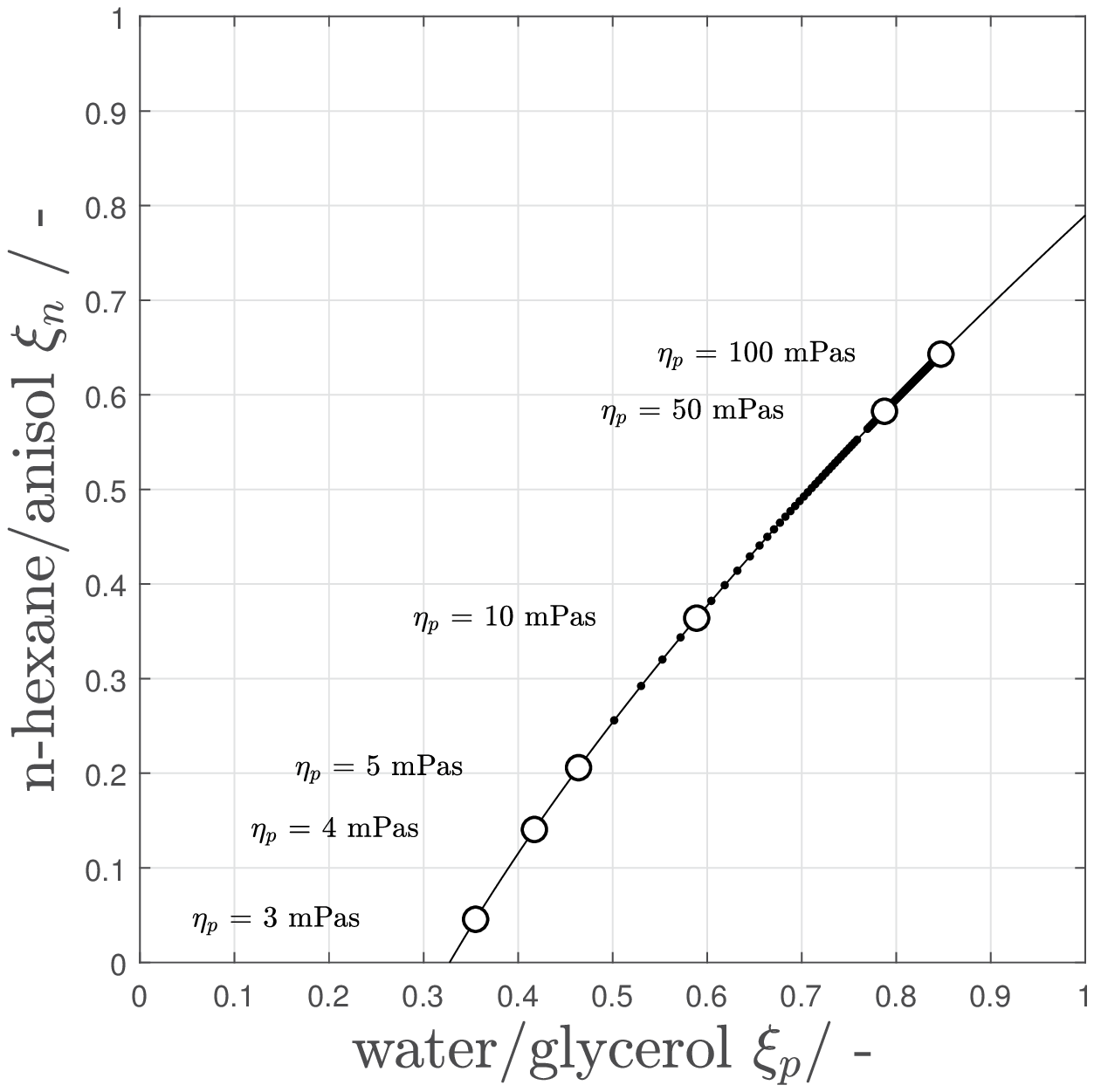}}%
	\qquad
	\subfloat[][]{\includegraphics[width=0.4\linewidth]{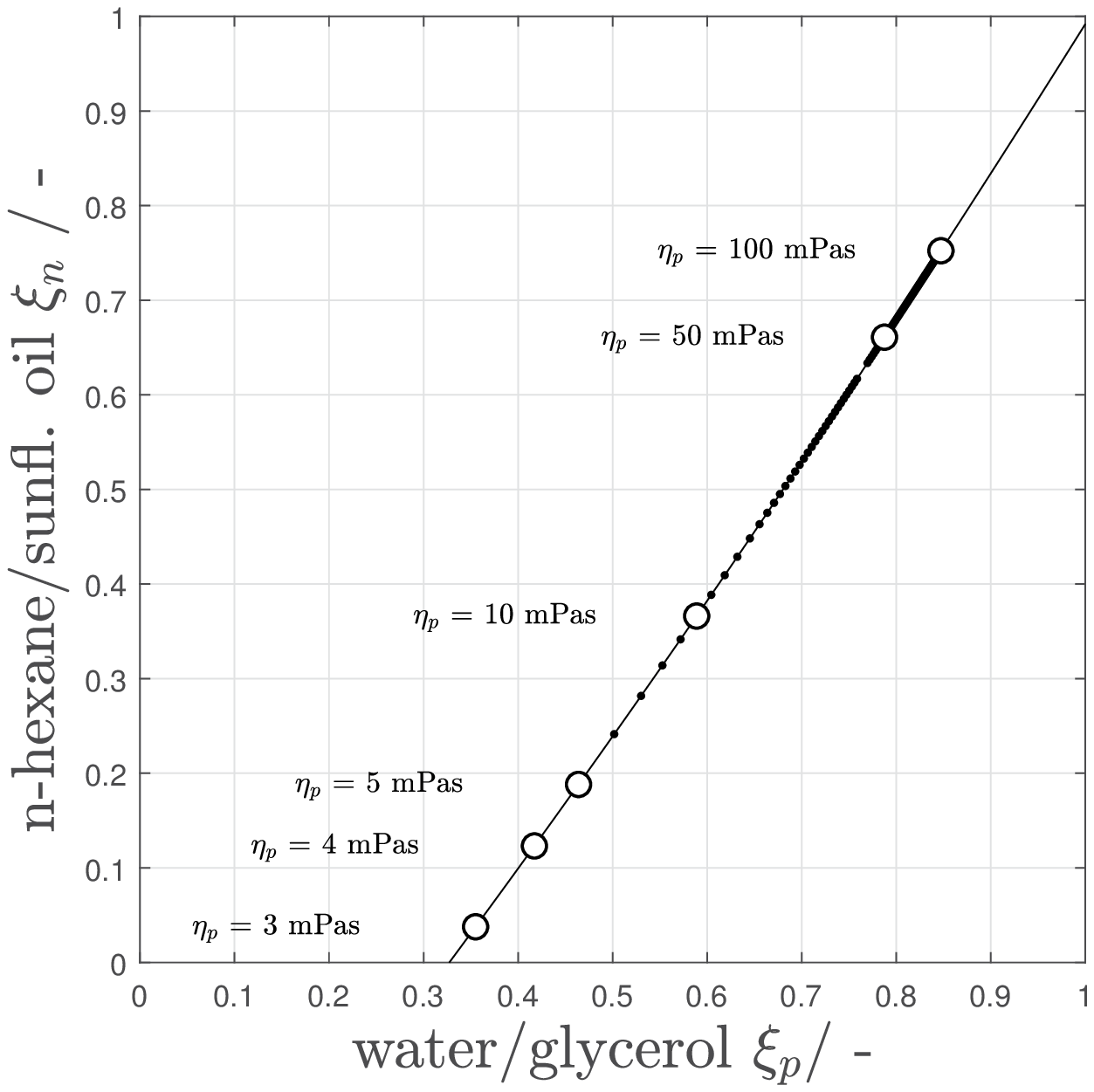}}%
	\caption{Polar continuous phase viscosity $\eta_p$ along the mass fractions for matched mixtures. a) water/glycerol - n-hexane/anisole b) water/glycerol - n-hexane/sunflower-oil. Large circles represent a logarithmic scale, small dots the interiorly intermediate steps. Low viscous continuous phase systems are not shown since the continuous viscosity does not change significantly over the RI-matched mass regions}
	\label{fig:doub-bin:matched-visco-p}       
\end{figure*}

In addition to the direct properties of the individual binary mixtures that have been discussed in the previous section, the properties of the coupled \textbf{RI-matched} material systems (double-binary mixtures) are discussed in this section. Since the double-binary system consists of four substances, the interfacial tension is influenced independently and possibly nonlinearly by the mass fractions of both binary mixtures. Thus to obtain a manageable experimental effort for the required measurements, the interrelations are linearized and the system is simplified:
For the use in RI-matched measurements, only the mass fractions of the matched solution need to be observed. As it is visible in Fig. \ref{fig:doub-bin:matched-ri} and \ref{fig:doub-bin:matched-visco-p} this simplifies the problem to two-dimensional problem. 

To retrieve the specific mass fractions of both phases for the matched case, the correlations for the RI of both phases are equated and a fit function is numerically retrieved. The behavior of the RI-matched double-binary systems is a combination of the binary mixtures. The results are shown for all double-binary mixture systems in Fig. \ref{fig:doub-bin:matched-ri}.

 The system water/DMSO-hexane/anisole (Fig. \ref{fig:doub-bin:matched-ri} (c)) shows the most prominent nonlinear behavior. This is caused by the larger increase of the RI of anisole for elevated anisole mass fractions and the nonlinearity of water/DMSO. For the system water/glycerol-n-hexane/anisole (Fig. \ref{fig:doub-bin:matched-ri} (a)) this behavior is as not prominent since the binary mixture water/glycerol in contrast to water/DMSO is nearly linear. As expected from the binary mixture's behavior, this also holds for the system water/glycerol-hexane/sunflower-oil (Fig. \ref{fig:doub-bin:matched-ri} (b)), while again caused by water/DMSO, for water/DMSO-n-hexane/sunflower-oil nonlinearities are present at higher DMSO mass fractions (Fig. \ref{fig:doub-bin:matched-ri} (d)). The correlation coefficients for the matched systems are shown in Tab. \ref{tab:doub-bin:corr_ri}.

\begin{table*}[htb]
	\centering
	\caption{Correlation coefficients for the mass fractions of RI-matched double-binary mixtures (range of validity for all correlations: $0.32 < \xi_p <1$) }
	\label{tab:doub-bin:corr_ri}       
	\begin{tabular}{l|r|r|r|r|r}
		\hline\noalign{\smallskip}
		\textbf{mixture} & $A_4$ & $A_3$ & $A_2$ & $A_1$ & $A_0$ \\
		\noalign{\smallskip}\hline\noalign{\smallskip}
		water/glycerol - hexane/anisole     & -4.599 & 12.318 & -12.348 & 6.632 & -1.237 \\
		water/glycerol - hexane/sunfl. oil  & 0      &   0    &   0.182 & 1.233 & -0.423 \\
		water/DMSO - hexane/anisole         & -3.623 & 9.629  & -10.027 & 5.908 & -1.098 \\
		water/DMSO -  hexane/sunfl. oil     & 0      & -0.631 & 0.835   & 1.236 & -0.453 \\
		\noalign{\smallskip}\hline
	\end{tabular}
\end{table*}   

Additionally to the refractive index, the densities, interfacial tension and viscosities for the different matched mass fraction have to be considered for the calculation of the desired dimensionless quantities ($Re$, $Ca$, $\lambda$). Every matched system consists of two binary mixtures with exactly one RI for the matched case and one specific mass fraction and viscosity for the polar as well as the disperse phase. Changing the mass fraction of one phase, the mass fraction of the other phase needs to be adjusted to keep the RI matching. This changes the viscosity ratio, as it also depends on the ratio of the binary mixtures of each phase. 

For the determination of the dimensionless quantities $Re$, $Ca$, $Oh$ of microscopic two-phase flows, especially the \textbf{viscosity} of the continuous phase is important, since it influences the dimensionless quantities the strongest. Therefore the viscosities for the continuous phase of the high viscous systems are shown (Fig. \ref{fig:doub-bin:matched-visco-p}). 

At first, the polar phase is depicted as the continuous phase, resulting in the graphs of Fig. \ref{fig:doub-bin:matched-visco-p}. The combinations with water/glycerol show a broad range of available continuous phase viscosities.

Considering the viscosities of both flow phases, the viscosity ratio $\lambda$ of the respective RI-matched material systems can be calculated. The results are shown in Fig. \ref{fig:doub-bin:lambda} separated for polar and nonpolar continuous phase:

\begin{figure*}[htb]
	\centering
	\subfloat[][]{\includegraphics[width=0.4\linewidth]{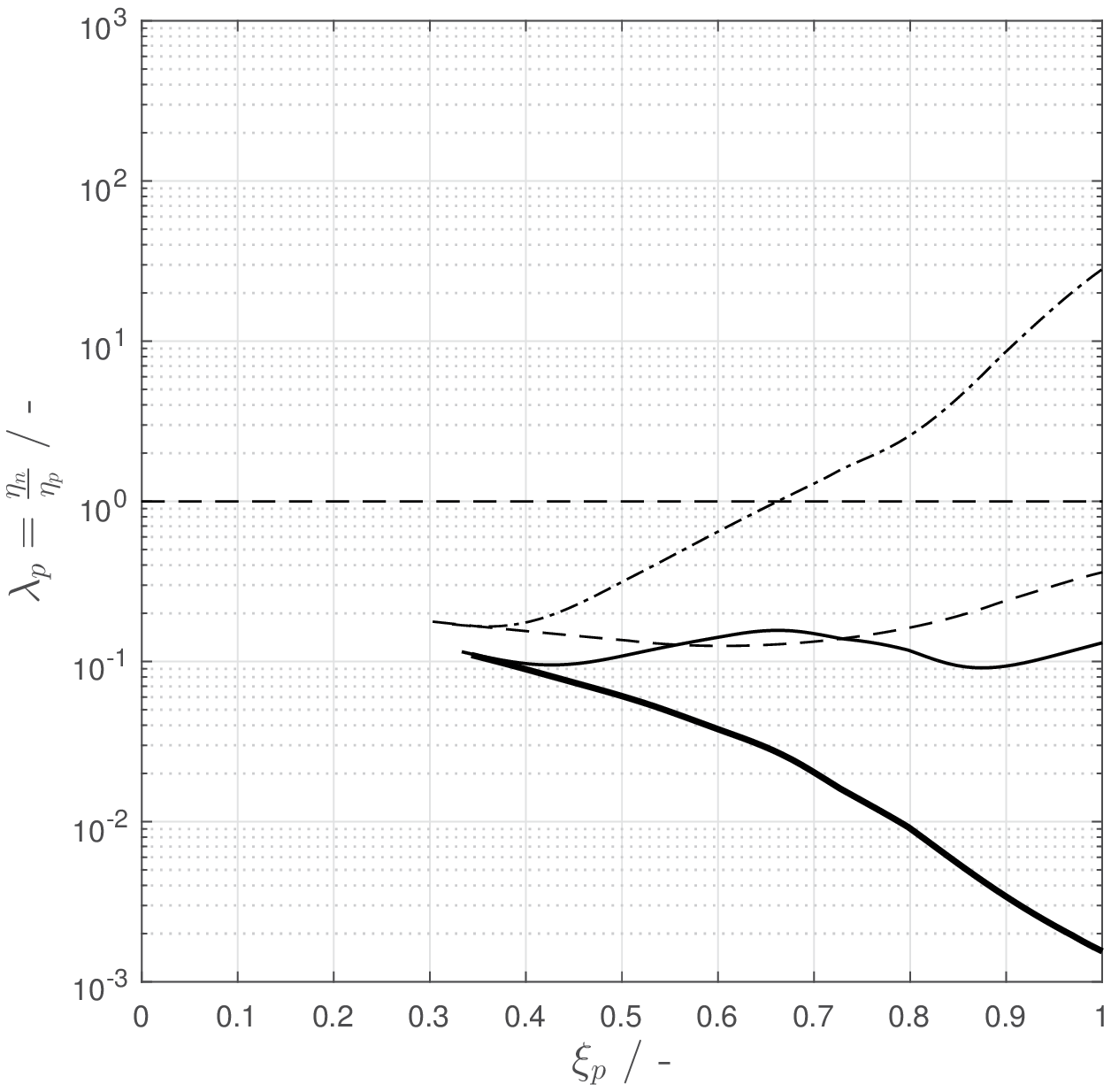}}%
	\qquad
	\subfloat[][]{\includegraphics[width=0.4\linewidth]{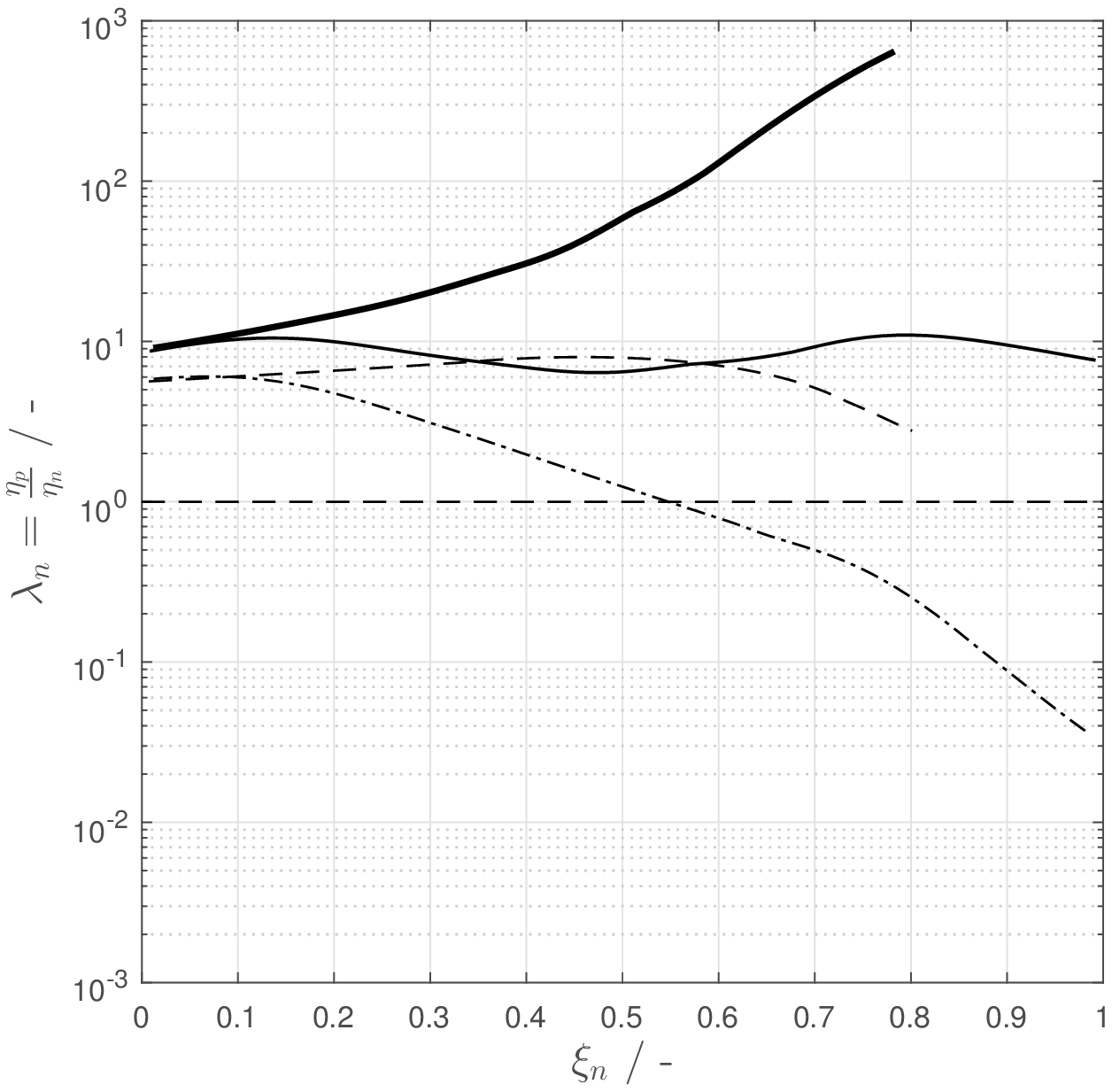}}%
	\caption{Viscosity ratios $\lambda_n$ and $\lambda_p$ for the different systems. System water/glycerol - n-hexane/anisole (bold solid line), system water/glycerol - n-hexane/sunflower-oil (thin solid line), system water/DMSO - n-hexane/anisole (dashed line), system  water/DMSO - n-hexane/sunflower-oil (dash-dotted line). a) continuous phase: polar b) continuous phase: nonpolar. }
	\label{fig:doub-bin:lambda}       
\end{figure*}

In Addition to the determination of the material properties that depend only on the binary mixtures themselves, also the interfacial tension needs to be described. Since it depends on the forces of the energetic state of the interfacial area, the determination for a quarternary fluid system is complex. For the following measurements, only the interfacial tension of the RI-matched case is of interest. Thus, we only determine values at the mass fractions of the RI-matched interfacial tensions to minimize the experimental effort.

\begin{figure*}[htb]
	\centering
	\subfloat[][]{\includegraphics[width=0.4\linewidth]{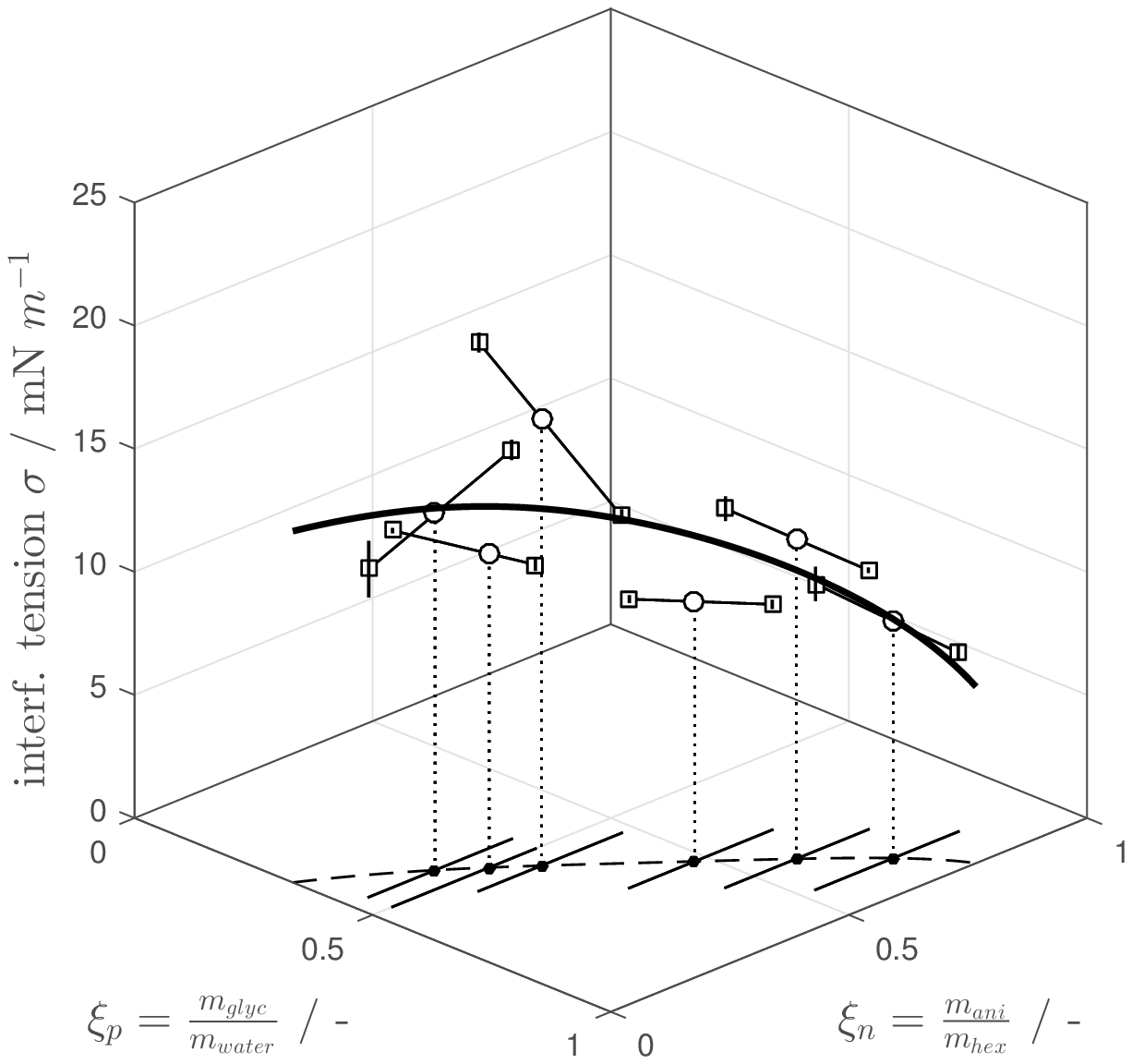}}%
	\qquad
	\subfloat[][]{\includegraphics[width=0.4\linewidth]{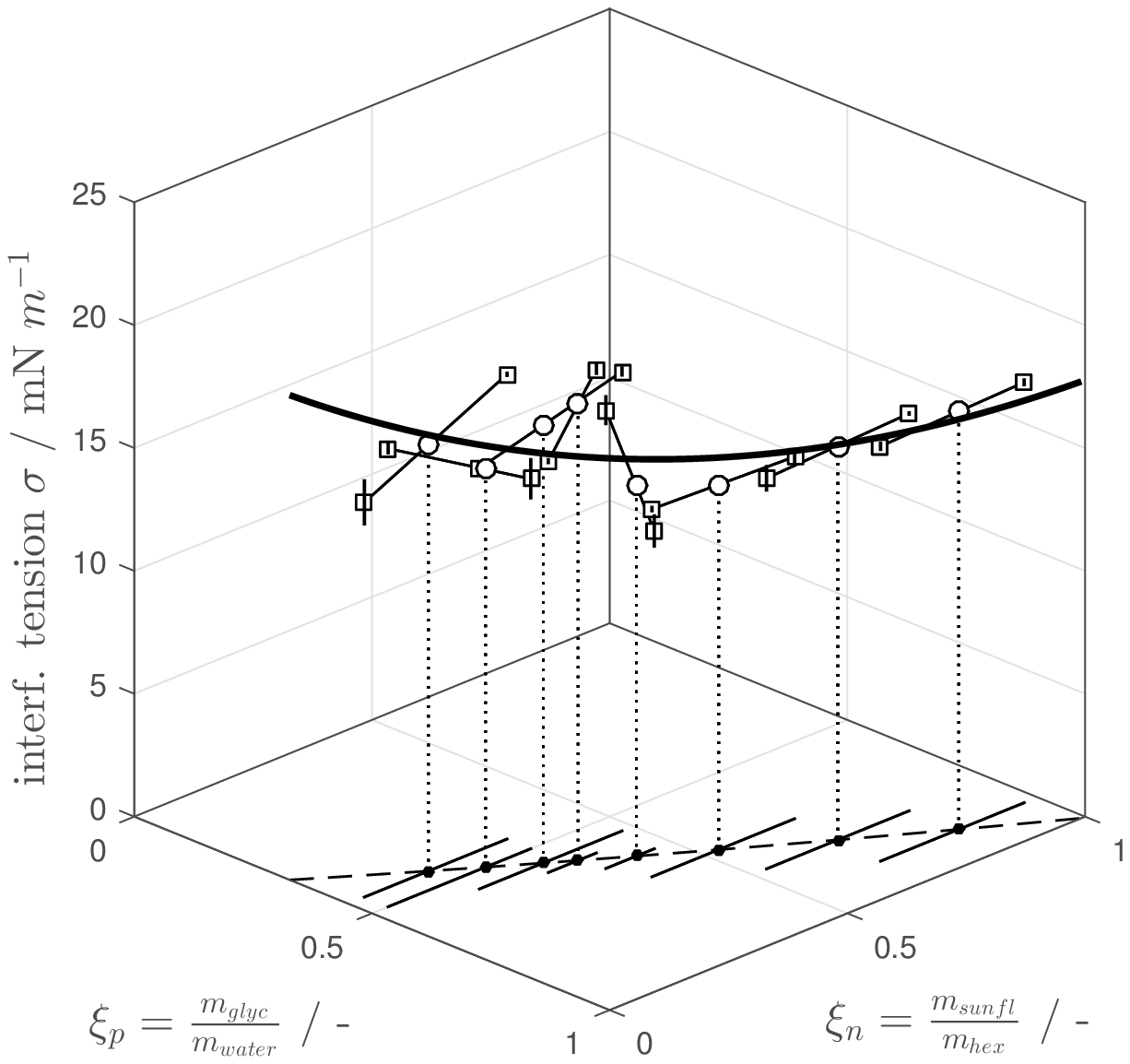}}%
	
	\subfloat[][]{\includegraphics[width=0.4\linewidth]{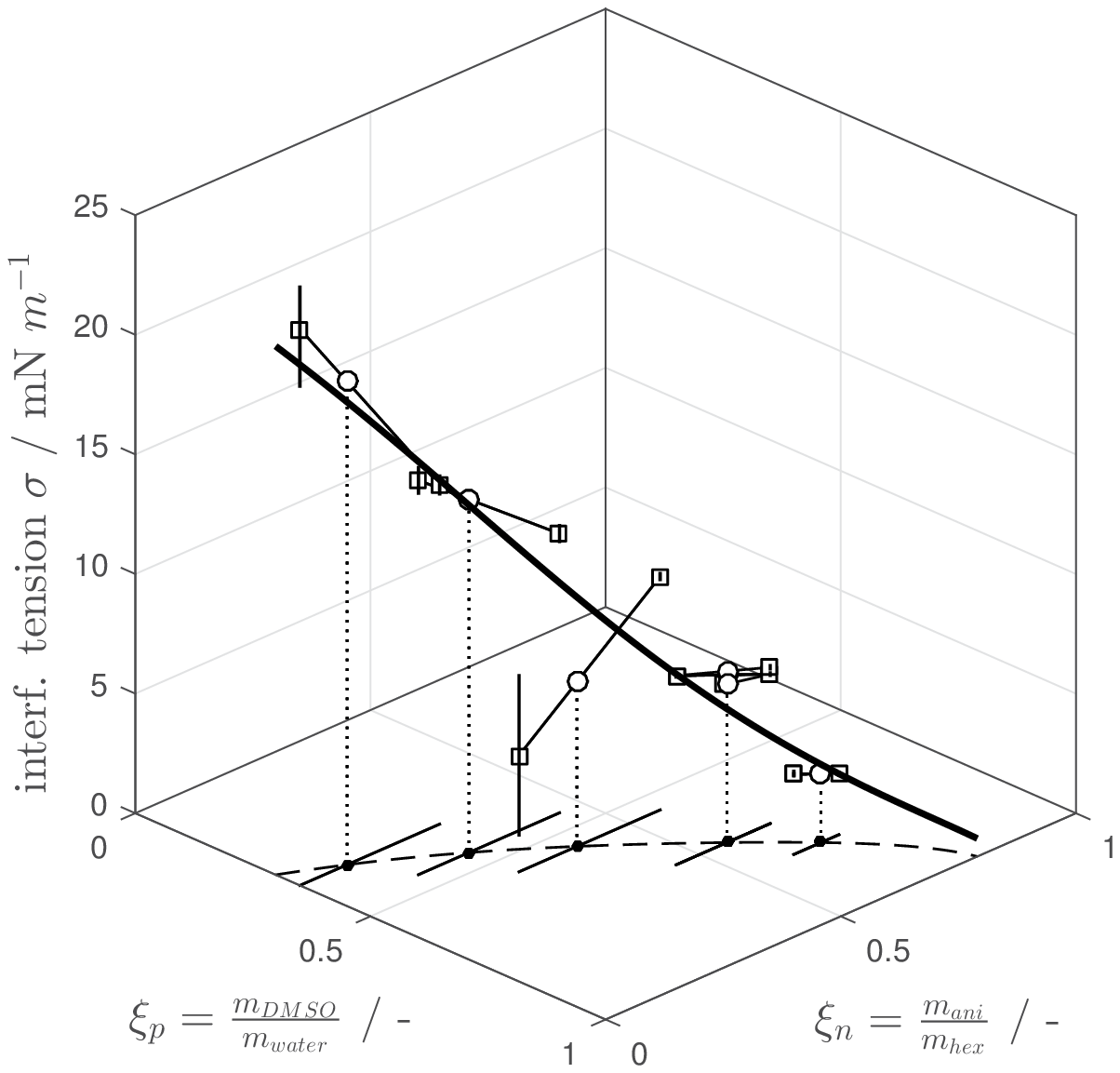}}%
	\qquad
	\subfloat[][]{\includegraphics[width=0.4\linewidth]{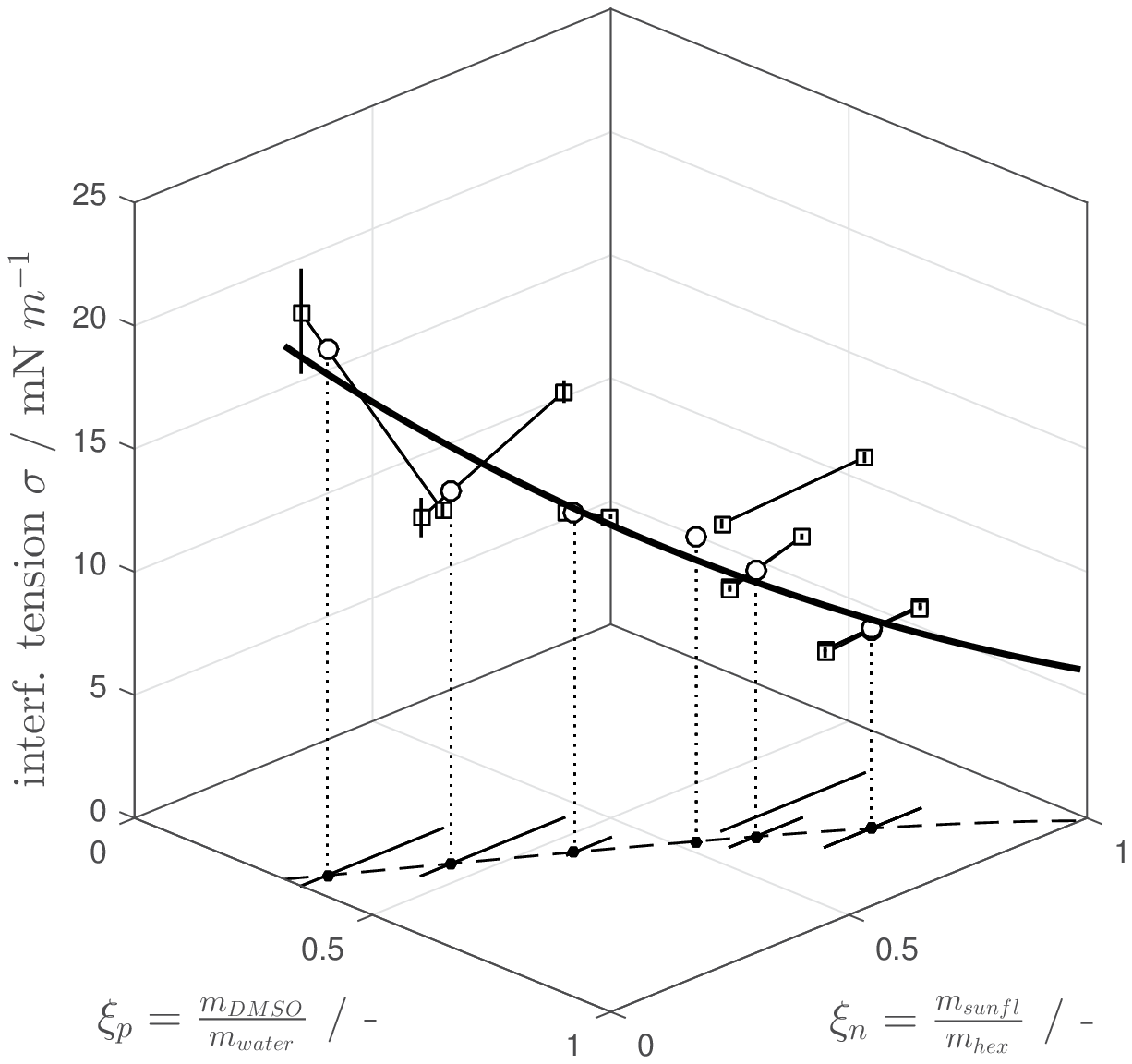}}%
	\caption{Interfacial tension $\sigma$ for the different systems. a) water/glycerol - n-hexane/anisole b) water/glycerol - n-hexane/sunflower-oil c) water/DMSO - n-hexane/anisole d) water/DMSO - n-hexane/sunflower-oil. Actual measurements (squares) are used to interpolate (bars) the interfacial tension at the matched phase composition (circles). The correlation for the matched interfacial tension is the correlated (bold line). Each measurement consists of 5 droplet formation cycles at different droplet formation times. Within each cycle, at least 5 droplets are formed to ensure reproducibility}
	\label{fig:doub-bin:matched-sigmas}       
\end{figure*}

Due to the measurement restrictions (see in Sec. \ref{sec:doub-bin:meas}), we perform the measurements at slightly detuned refractive indices and linearly interpolate the interfacial tension for the matched case between the results. The measurement data, as well as the derived correlations, are shown in Fig. \ref{fig:doub-bin:matched-sigmas}.

The interfacial tension between the RI-matched liquids exhibits different behavior for each system. The system water/glycerol- hexane/anisole (Fig. \ref{fig:doub-bin:matched-sigmas} a)) shows an unpredictable interfacial tension development for rising mass fractions of glycerol or anisole. The interpolated values of the single measurements deviate around the correlation function. This is a sign for strong nonlinear interrelations between the four substances. Probably caused by weaker intermolecular forces between glycerol and anisole depending on the composition of the mixture. The system water/glycerol - hexane/sunflower-oil shows a minimum for equivalent fractions of all four substances ($\xi_p \approx 0.5$, $\xi_n \approx 0.5$), while the interfacial tension increases for higher amounts of glycerol or sunflower oil (Fig. \ref{fig:doub-bin:matched-sigmas} b)).

The systems involving water/DMSO expose a decreasing interfacial tension with decreasing mass-fraction of n-hexane. Especially the water/DMSO - hexane/anisole shows a very low interfacial tension for lower mass fractions of hexane (Fig. \ref{fig:doub-bin:matched-sigmas} c)), which is caused by a diminished structural difference to hexane. Water/DMSO - n-hexane/sunflower-oil shows a similar decrease of the interfacial tension (Fig. \ref{fig:doub-bin:matched-sigmas} d)). 

The most significant deviations (Fig. \ref{fig:doub-bin:matched-sigmas} a) and b)) are situated in a range of 2\,mPas. Please note, that the derived fitting function represents an averaging attempt to describe the complex 4-material-system. Therefore, a deviation between the correlation and the interpolated data of single measurements can be caused by either the linearization or confined effects at distinct mass fractions as well as measurement errors. Since the results of the single measurements are based on at least 20 independent droplets (5 droplets x 4 formation times), we consider possible measurement errors to attribute from diminutive contaminations of the volume tensiometer. The correlation coefficients for the interfacial tension for all four systems are shown in Tab. \ref{tab:doub-bin:corr_intf}.

\begin{table}[h]
	\caption{Correlations coefficients for interfacial tension (range of validity $0.32 < \xi_p <1.00$)}
	\label{tab:doub-bin:corr_intf}       
	\begin{tabular}{l|r|r|r}
		\hline\noalign{\smallskip}
		\textbf{mixture} &$A_2$ & $A_1$ & $A_0$ \\
		\noalign{\smallskip}\hline\noalign{\smallskip}
		wt./glycerol - hex./anisole       & -26.002 & -24.007 & 9.164 \\
		wt./glycerol - hex./sunfl. oil & 24.029 & -34.799 & 28.538 \\
		wt./DMSO - hex./anisole           &   35.627 & -76.700 & 41.824 \\
		wt./DMSO - hex./sunfl. oil     &  22.616  & -52.398 & 35.929 \\
		\noalign{\smallskip}\hline
	\end{tabular}
\end{table}  

\section{Discussion}\label{sec:doub-bin:disc}
Based on the material properties of the binary mixtures as well as the double-binary mixtures, which were experimentally determined and successfully modeled with polynomial correlations, we describe the features of the double-binary approach in this section and deliver a proof-of-principle.

\subsection{Features and limits of the double-binary approach}\label{sec:doub-bin:possib}
\begin{figure}[h]
	\centering
	\includegraphics[width=0.8\linewidth]{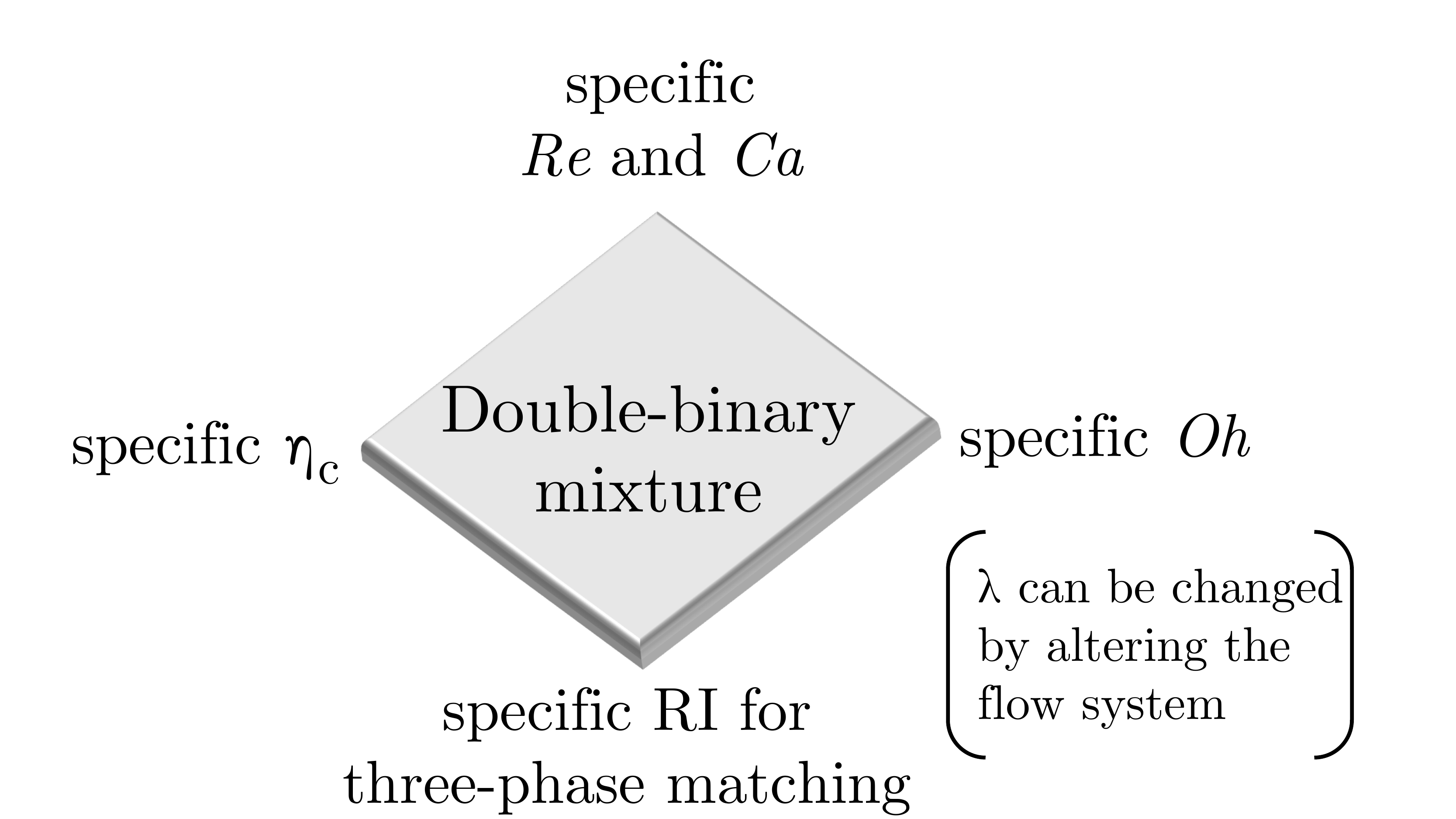}
	\caption{Features of the double-binary fluid system. Depending on the case of application, $Re$ and $Ca$ can be adjusted individually. Alternatively, a defined continuous phase viscosity $\eta_c$, a defined RI or a defined $Oh$ can be chosen. The latter three can only be addressed at a defined $Re$ or $Ca$, while the other dimensionless quantity is fixed}
	\label{fig:doub-bin:possibilities}
\end{figure}

The proposed double-binary mixture approach allows tuning the material system of e.g. microscopic Taylor flows with an additional degree of freedom. A two-phase flow can now be RI-matched at a specific individually set of $Re$ and $Ca$ individually. This is done via a variation of the mass-fraction of both phases $\xi_{c}$, $\xi_{d}$ and the superficial velocity $u_0$. Alternatively, a set of fixed $Ca$/$Re$ ratios ($\sim Oh$) at distinct RI can be adjusted, allowing to match the system to a specific reactor material. In this way even complex structures or three phase flows can be observed once the flow is matched to the reactor material. Additionally, the flow can be applied at a specific viscosity of the continuous phase (if fluids with a distinct viscosity should be modeled). A graphical representation of the capabilities is given in Fig. \ref{fig:doub-bin:possibilities}.

\begin{figure*}[htb]
	\centering
	\subfloat[][]{\includegraphics[width=0.4\linewidth]{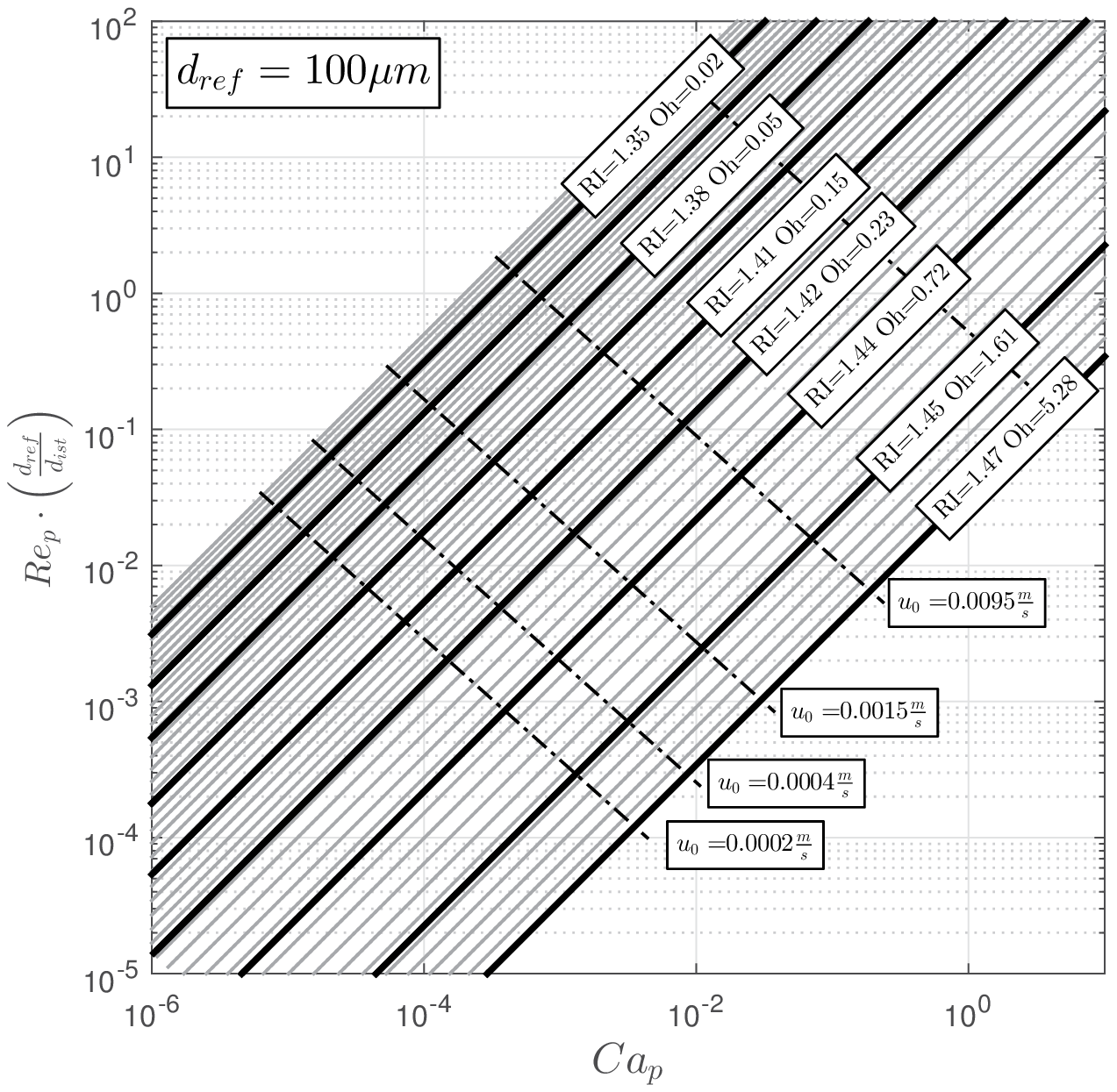}}%
	\qquad
	\subfloat[][]{\includegraphics[width=0.4\linewidth]{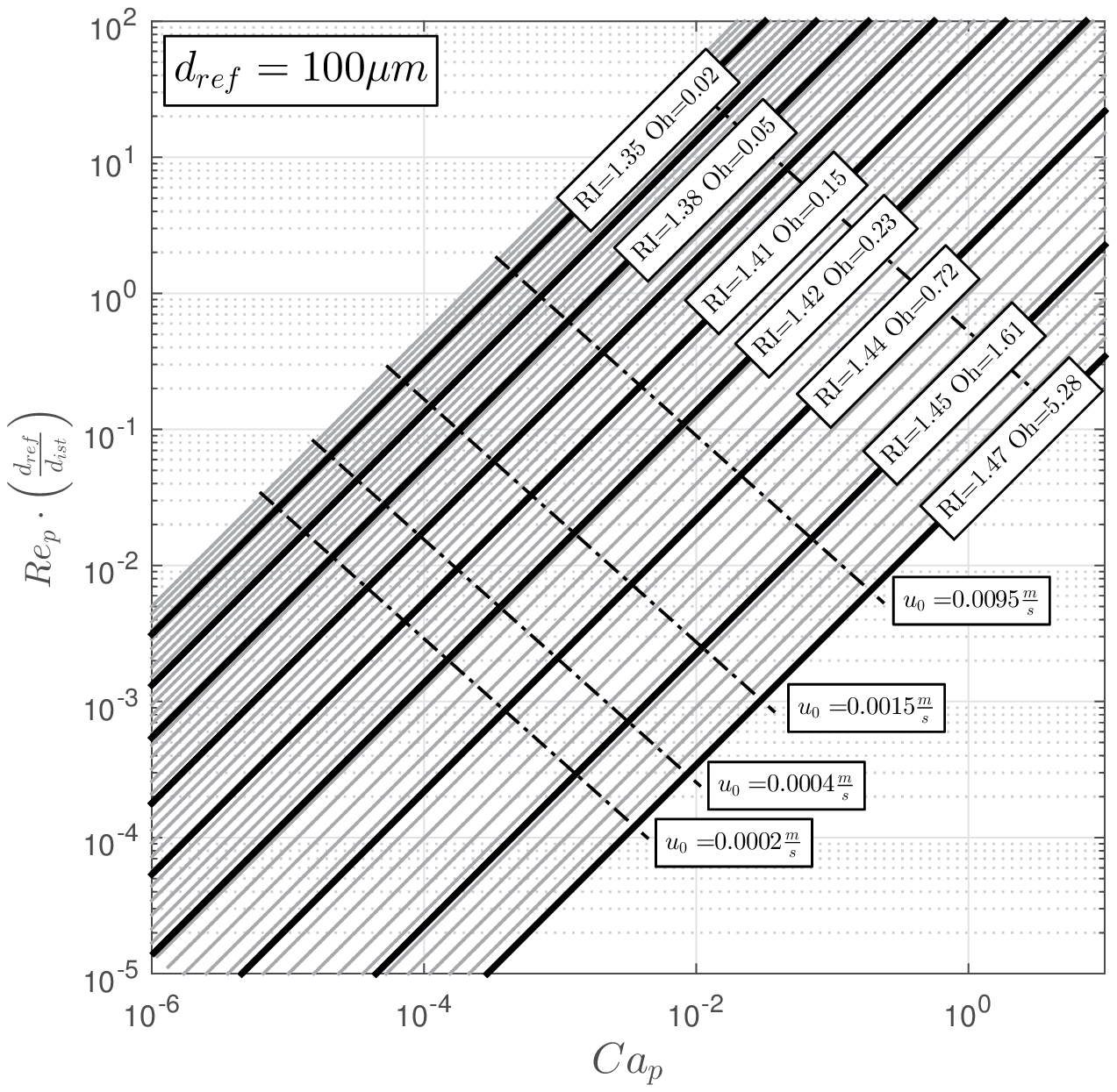}}%

	\subfloat[][]{\includegraphics[width=0.4\linewidth]{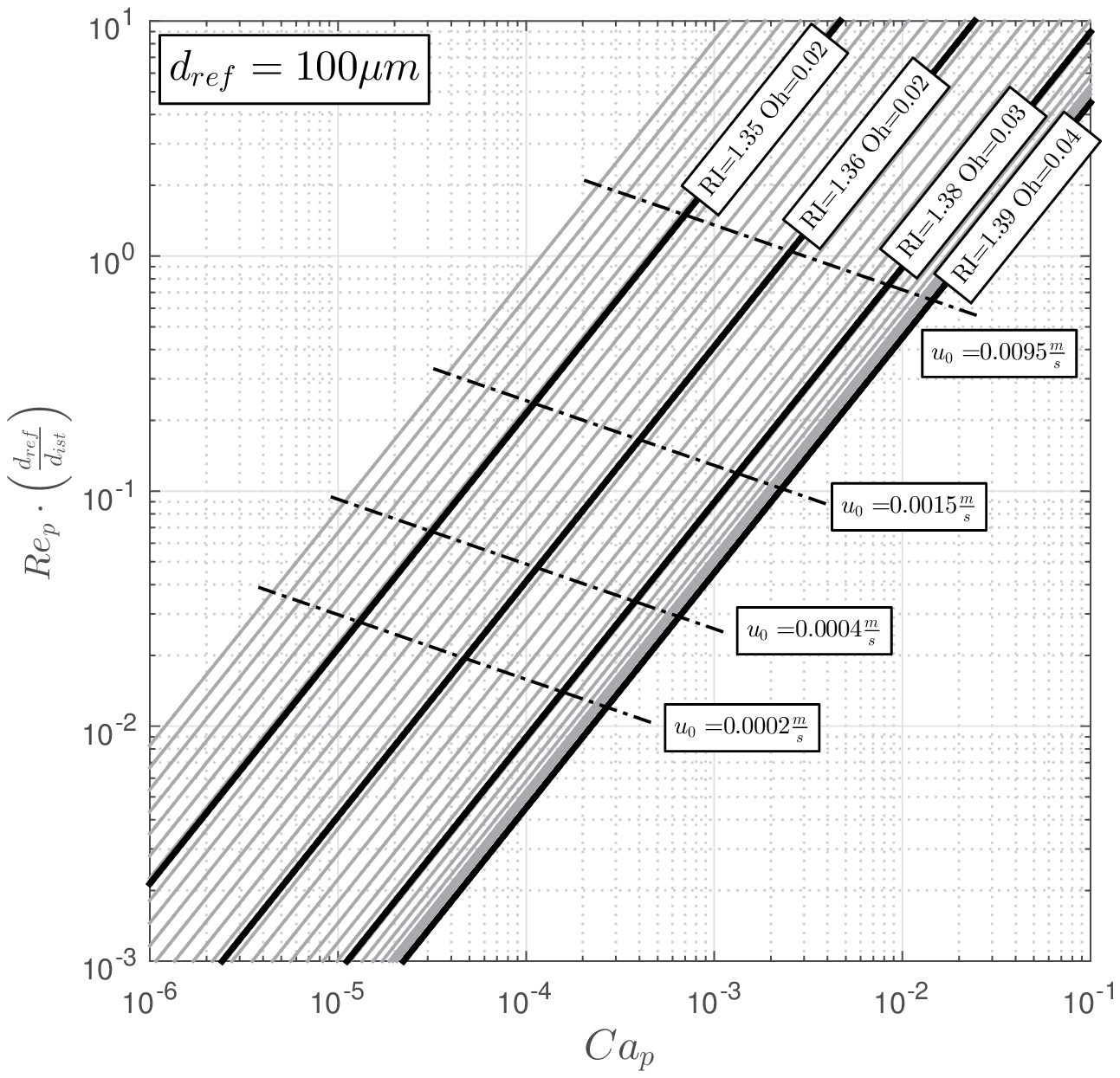}}%
	\qquad
	\subfloat[][]{\includegraphics[width=0.4\linewidth]{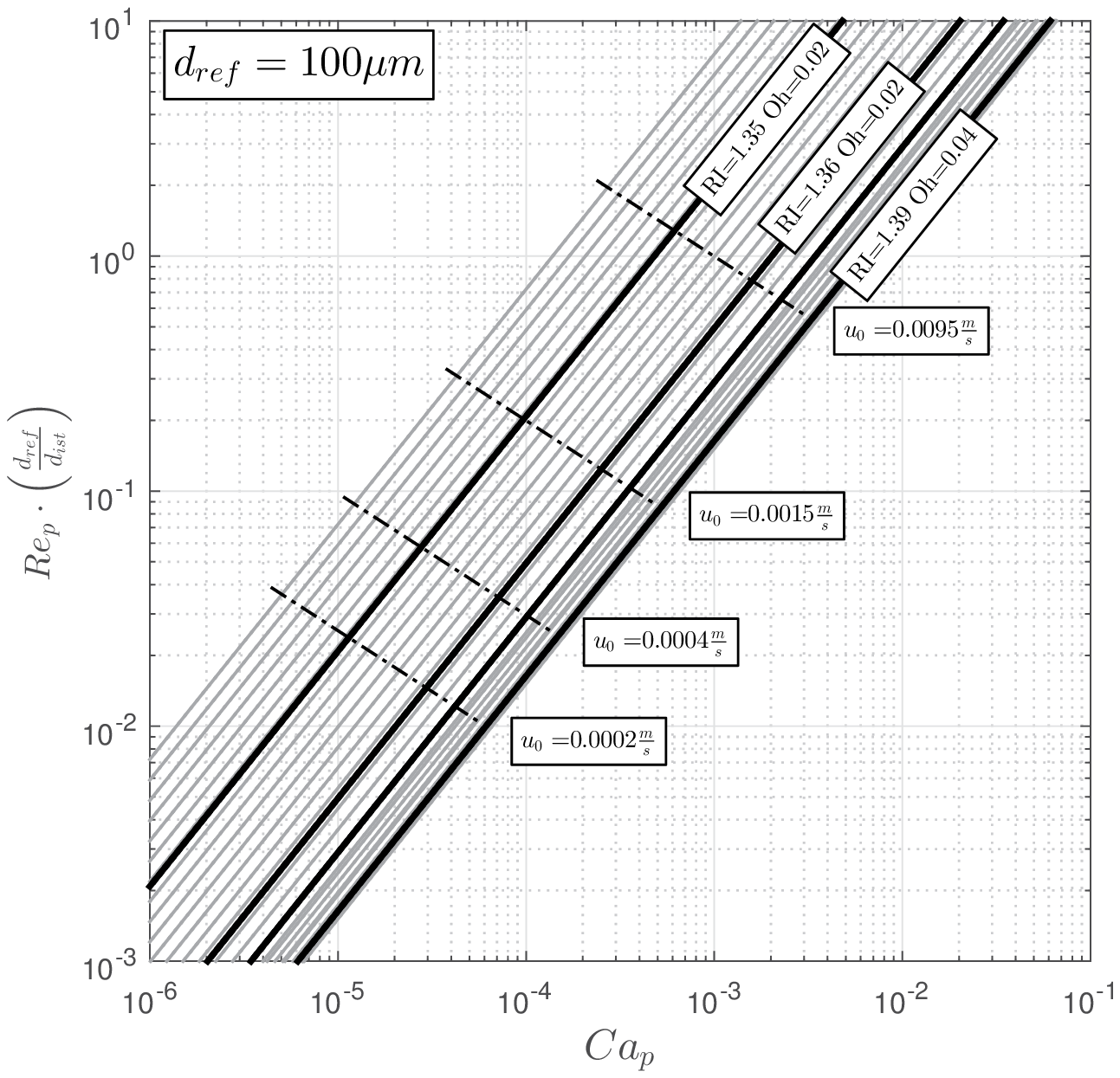}}%
	\caption{Nomograms for reachable $Ca$ and $Re$ numbers for a polar continuous phase. a) water/glycerol - n-hexane/anisole b) water/glycerol - n-hexane/sunflower-oil c) water/DMSO - n-hexane/anisole d) water/DMSO - n-hexane/sunflower-oil}
	\label{fig:doub-bin:matched-nomos}       
\end{figure*}

Based on the conducted measurements and literature data, the proposed double-binary mixture systems are sufficiently described to calculate the relevant dimensionless quantities $Ca$ and $Re$ for all accessible mass fractions and superficial velocities. These  latter two quantities can be calculated by the MATLAB program we supply with the supplementary material. Alternatively, they mass fraction and the superficial velocity can be retrieved graphically from the nomograms presented in Fig. \ref{fig:doub-bin:matched-nomos}. The desired $Ca$ and $Re$ are chosen on the according axis and the corresponding matched $Oh$ and RI of both phases, as well as the superficial velocity, can be received. The RI determines the mass fractions of both phases $\xi$ (Fig. \ref{fig:doub-bin:matched-ri}) and the associated viscosity ratio (Fig. \ref{fig:doub-bin:lambda}). Via the continuous phase viscosity (Fig. \ref{fig:doub-bin:matched-visco-p}) and $\lambda$ the disperse phase viscosity can be determined. Generally, the systems with a high viscous continuous phase (Fig. \ref{fig:doub-bin:matched-nomos} a) + b)) cover a larger range of possible $Ca$ and $Re$, while for the lower viscous continuous phase the parameter range is smaller.

\subsection{Proof of Principle}
The practicability of the proposed approach for optical measurements and especially the possibility to reach the desired $Re$ and $Ca$ independently while a matched refractive index is proven in an experimental approach. A microscopic Taylor flow is established at two different $Re$ numbers while $Ca$ is held constant. The experimental setup to record the PIV raw-images is described in the appendix. 

\nomenclature[U]{$tr$}{tracer particle}

\begin{figure}[htb]
\centering
\includegraphics[width=0.3\linewidth]{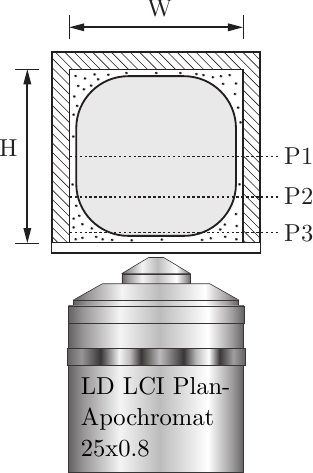}
	\label{fig:doub-bin:pop_obje}       
	\caption{Cross-section of the microchannel with the measurement planes for proof of principle}
\end{figure}

The capability of the double-binary approach and the quality of refractive index matching is evaluated in three measurement planes at two $Re$ numbers at a steady $Ca$. Measurements in the symmetry plane (channel center plane), the channel top wall and at an intermediate plane at 0.75 of the channel height are carried out (Fig. \ref{fig:doub-bin:pop_obje}). In the symmetry plane at the half channel height, distortions arising from non-ideal refractive index matching accumulate and can be quantified. The measurements at the channel top allow classifying, if the flow through the gutters can be measured.

\begin{figure*}[htb]
	\centering
	\subfloat[][]{\includegraphics[width=0.42\linewidth]{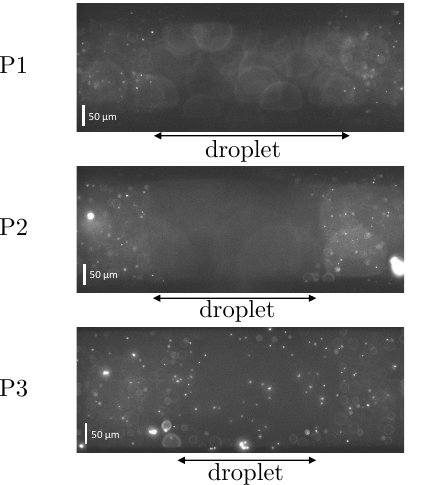}}%
	\qquad
	\subfloat[][]{\includegraphics[width=0.42\linewidth]{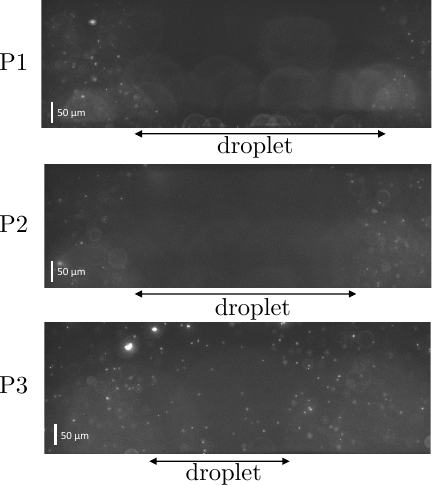}}%
	\caption{Exemplary images of a Taylor-flow at two different $Re$ and steady $Ca$. Water/glycerol is used as the continuous phase and seeded with fluorescent polystyrene particles, n-hexane/sunflower-oil as the disperse phase. Within all measurements no interfacial area is distinguishable, no reflection or distortions of the particles are apparent. a) $Ca$ = 0.005, $Re$ = 2.14, RI = 1.3838 $\xi_p$ = 0.391 $\xi_n$ = 0.101 b) $Ca$ = 0.005, $Re$ = 0.93, RI = 1.3958 $\xi_p$ = 0.476 $\xi_n$ = 0.221 }
	\label{fig:doub-bin:pop2}       
\end{figure*}

Within the \textmu PIV raw images, the droplet interface is not visible. The raw-images in Fig. \ref{fig:doub-bin:pop2} show clearly that the droplet length decreases when the focal plane is moved towards the channel top. This is caused by the curvature of the droplet interface. Only the tracer particles in the focal plane are displayed sharp, while particles on other planes out of focus introduce a blurry signal.  Thus, the correlations for the refractive index as well as the solver algorithm works well. 

Taylor flows were established and recorded without optical distortion. Thus, the proposed double-binary mixture method works well. 

\section{Conclusion}\label{sec:doub-bin:concl}
Within this study, we present and successfully validate a new approach using double-binary mixtures for both immiscible flow phases to establish a refractive index matched microscopic multiphase flows. In comparison to classical mono-binary mixture approaches, $Re$ and $Ca$ can be addressed individually in a material restricted parameter set, since viscosity and interfacial tension (and thus the different flow forces) do not change in the same order of magnitude if the mass fractions of the phases' binary mixtures are varied. Alternatively, multiphase flows at different $Oh$ at a fixed $RI$ can be established to match the reactor material (e.g. simulating three-phase flows via monolith).

Additionally, we introduce two binary mixtures for a polar and nonpolar phase to enable the investigation of e.g., different viscosity ratios $\lambda$ or flow systems with simultaneously high or low viscosities are possible too.

Measurements for the relevant material properties (densities, viscosities, refractive index) of the binary mixtures as well as the RI-matched compositions (interfacial tensions) are carried out and compared to literature data to characterize the system. The task to establish  multiphase flows at specific $Re$ and $Ca$ independently is identified as an optimization problem and the material properties as well as the interfacial tension are successfully described with correlations that allow the use of a solver algorithm.

In first measurements, the capability to establish refractive index matched Taylor flows at freely chosen $Ca$ and $Re$ is proven. This proof of principle is successfully conducted using recorded \textmu PIV raw images. With the proposed double-binary RIM-approach the specific influences on the local velocity of droplets can now be independently examined via optical flow visualization techniques (PTV, PIV).
 
 \section*{Appendix}
\begin{figure*}[htb] 
	\centering
	\includegraphics[width=0.7\linewidth]{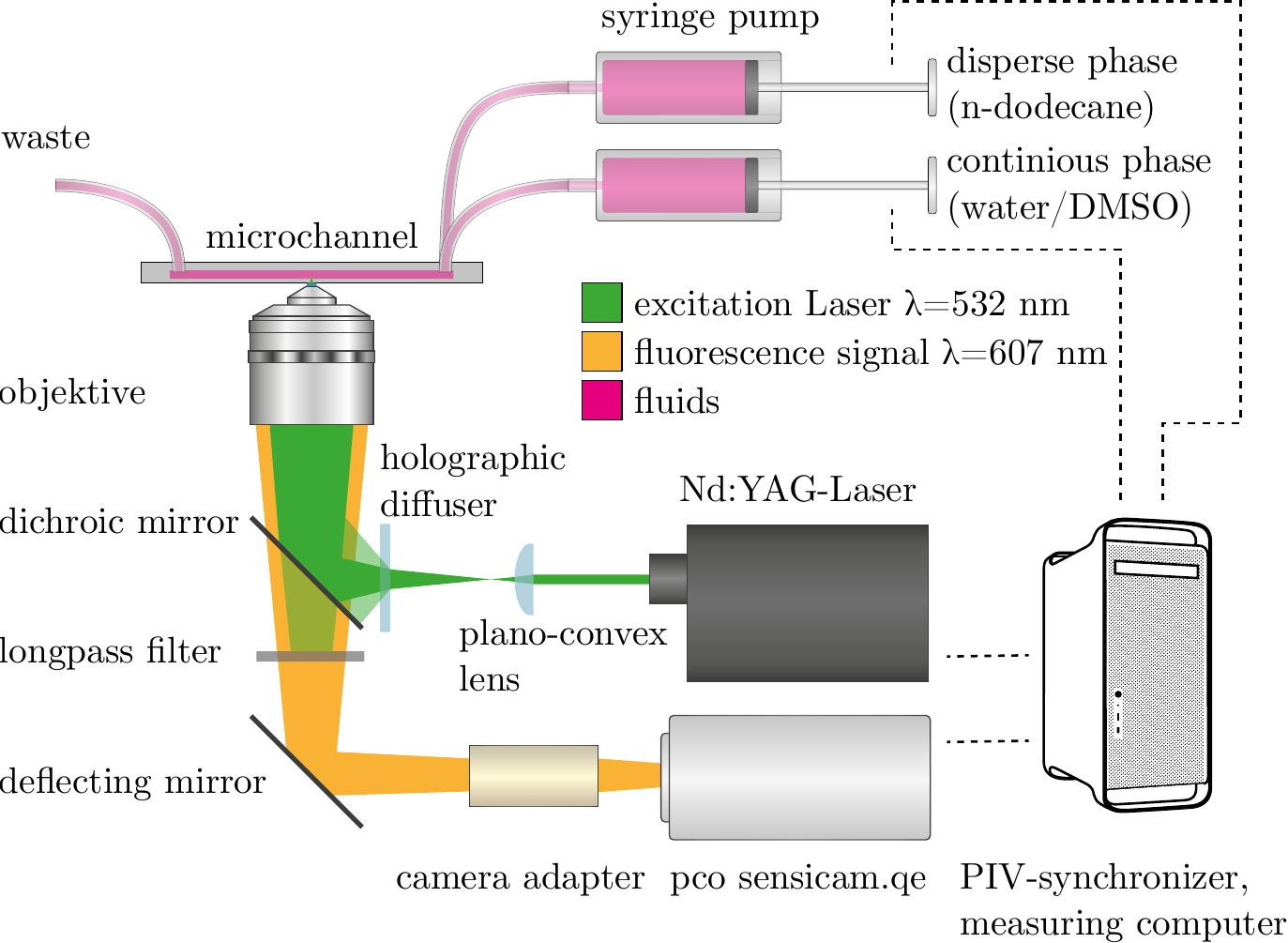}
	\caption{Proof of principle of the double-binary mixture approach. Scheme of the experimental design for \textmu PIV experiments}
	\label{fig:doub-bin:pop}
\end{figure*}

The experimental design follows Fig. \ref{fig:doub-bin:pop}: A  Si-microchannel (manufactured by IMSAS Bremen) with 198 \textmu m nearly rectangular cross-sectional area is located on a Zeiss LSM-210 inverted microscope with a motorized nosepiece and two-axis stage for precision movement. The microscope is controlled via a self-written LabView program using serial port communication.  

The microscope is equipped with a Zeiss LD LCI Plan Apochromat 25x/0.8 objective to provide a high spatial resolution at a small depth of field ($DOF = 1.034 \mu m$) to receive a high resolution in the z-direction. A pulsed Nd:YAG laser (\textit{New Wave Research Solo-PIV III}) with a 15 Hz repetition rate, 50 mJ pulse intensity and a pulse length of 3 ns - 5 ns with frequency doubling (wavelength: $\lambda_{ex}$ = 532 nm) serves as a light source for the measurement. Images of the flow are acquired using an active-cooled high quantum-efficient \textit{PCO.sensicam qe 670 LD 3078} double CCD-camera with an acquisition rate of 4 Hz at a resolution of 1376 px x 1040 px. The continuous phase is seeded with particles of 1.61 \textmu m particles. The particles are coated with FluoRed as fluorescence dye (excitation peak 530 \textmu m, emission peak 607 \textmu m) and are dispersed in the water/glycerol phase using an ultrasonic bath for 15 min at 20\textdegree C.  

Two syringe pumps (\textit{Dolomite Mitos Duo XS}) supply a steady flow such, that the volume flow of both phases can be individually controlled. The excitation light from the pulsed Nd:YAG laser is guided into the \textit{Zeiss LSM210} microscope and the laser is widened with a convex lens of short focal length. The illumination is additionally averaged with a holographic diffuser.

A dichroic mirror separates the green excitation light from the red fluorescence light and directs it to the camera. To further improve the cutoff, an additional long-pass filter is mounted in the light-path  to shield the camera from laser light (Fig. \ref{fig:doub-bin:pop}). A timing unit synchronizes the laser and the camera. 

An overview of the experimental and optical parameters is given in Tab. \ref{tab:doub-bin:pop-param}.

\begin{table}[htb]
	\caption{Experimental and optical parameters of the experimental design}
	\label{tab:doub-bin:pop-param}       
	\begin{tabular}{l|r|r}
		\hline\noalign{\smallskip}
		\textbf{property} &value & unit \\
		\noalign{\smallskip}\hline\noalign{\smallskip}
		channel width $W$      & 198 & \textmu m \\
		channel height $H$ 		& 192 & \textmu m \\
		excitation wavelength $\lambda_{ex}$ 		& 532 & nm \\
		tracer particle diameter $d_{tr}$     &  1.6  & \textmu m \\
		tracer particle CV      &  2.3  & \% \\
	    objective magnification           &   25 & -\\
		objective NA     &  0.8  & - \\
		objective DOF     &  0.859  & \textmu m \\
		particle excitation wavelength peak  	& 530 & nm\\
		particle emission wavelength peak 		& 607 & nm\\
		dichroitic mirror cutoff                &  552  & nm \\
		longpass filter cutoff                  &  550  & nm \\
		\noalign{\smallskip}\hline
	\end{tabular}
\end{table}  

\bibliography{references}   

\begin{thebibliography}{42}
\providecommand{\natexlab}[1]{#1}
\providecommand{\url}[1]{{#1}}
\providecommand{\urlprefix}{URL }
\expandafter\ifx\csname urlstyle\endcsname\relax
  \providecommand{\doi}[1]{DOI~\discretionary{}{}{}#1}\else
  \providecommand{\doi}{DOI~\discretionary{}{}{}\begingroup
  \urlstyle{rm}\Url}\fi
\providecommand{\eprint}[2][]{\url{#2}}

\bibitem[{Ahmed et~al(2018)Ahmed, Akram, Bule, and Iqbal}]{Ahmed.2018}
Ahmed I, Akram Z, Bule M, Iqbal H (2018) Advancements and potential
  applications of microfluidic approaches---a review. Chemosensors 6(4):46,
  \doi{10.3390/chemosensors6040046}

\bibitem[{Al-Jimaz et~al(2005)Al-Jimaz, Al-Kandary, Abdul-latif, and
  Al-Zanki}]{AlJimaz.2005}
Al-Jimaz AS, Al-Kandary JA, Abdul-latif AHM, Al-Zanki AM (2005) Physical
  properties of {anisole+n-alkanes} at temperatures between (293.15 and 303.15)
  k. The Journal of Chemical Thermodynamics 37(7):631--642,
  \doi{10.1016/j.jct.2004.09.021}

\bibitem[{Ara{\'u}jo et~al(2012)Ara{\'u}jo, Miranda, Pinto, and
  Campos}]{Araujo.2012}
Ara{\'u}jo J, Miranda JM, Pinto A, Campos J (2012) Wide-ranging survey on the
  laminar flow of individual taylor bubbles rising through stagnant newtonian
  liquids. International Journal of Multiphase Flow 43:131--148,
  \doi{10.1016/j.ijmultiphaseflow.2012.03.007}

\bibitem[{Brindise et~al(2018)Brindise, Busse, and Vlachos}]{Brindise.2018}
Brindise MC, Busse MM, Vlachos PP (2018) Density- and viscosity-matched
  newtonian and non-newtonian blood-analog solutions with pdms refractive
  index. Experiments in Fluids 59(11):38, \doi{10.1007/s00348-018-2629-6}

\bibitem[{Budwig(1994)}]{Budwig.1994}
Budwig R (1994) Refractive index matching methods for liquid flow
  investigations. Experiments in Fluids 17(5):350--355,
  \doi{10.1007/BF01874416}

\bibitem[{Cadillon et~al(2016)Cadillon, Saksena, and
  Pearlstein}]{Cadillon.2016}
Cadillon J, Saksena R, Pearlstein AJ (2016) Transparent, immiscible, surrogate
  liquids with matchable refractive indexes: Increased range of density and
  viscosity ratios. Physics of Fluids 28(12):127,102, \doi{10.1063/1.4968512}

\bibitem[{Chen et~al(2014)Chen, Li, Huang, Xie, Mai, Wang, Nguyen, and
  Huang}]{Chen.2014}
Chen Y, Li P, Huang PH, Xie Y, Mai JD, Wang L, Nguyen NT, Huang TJ (2014) Rare
  cell isolation and analysis in microfluidics. Lab on a chip 14(4):626--645,
  \doi{10.1039/c3lc90136j}

\bibitem[{Chou et~al(2015)Chou, Lee, Yang, Huang, and Lin}]{Chou.2015}
Chou WL, Lee PY, Yang CL, Huang WY, Lin YS (2015) Recent advances in
  applications of droplet microfluidics. Micromachines 6(9):1249--1271,
  \doi{10.3390/mi6091249}

\bibitem[{Clausell-Tormos et~al(2008)Clausell-Tormos, Lieber, Baret, El-Harrak,
  Miller, Frenz, Blouwolff, Humphry, Köster, Duan, Holtze, Weitz, Griffiths,
  and Merten}]{ClausellTormos.2008}
Clausell-Tormos J, Lieber D, Baret JC, El-Harrak A, Miller OJ, Frenz L,
  Blouwolff J, Humphry KJ, Köster S, Duan H, Holtze C, Weitz DA, Griffiths AD,
  Merten CA (2008) Droplet-based microfluidic platforms for the encapsulation
  and screening of mammalian cells and multicellular organisms. Chemistry \&
  Biology 15(5):427 -- 437,
  \doi{https://doi.org/10.1016/j.chembiol.2008.04.004},
  \urlprefix\url{http://www.sciencedirect.com/science/article/pii/S1074552108001506}

\bibitem[{Cl{\'e}ment et~al(2018)Cl{\'e}ment, Guillemain, McCleney, and
  Bardet}]{Clement.2018}
Cl{\'e}ment SA, Guillemain A, McCleney AB, Bardet PM (2018) Options for
  refractive index and viscosity matching to study variable density flows.
  Experiments in Fluids 59(2):434, \doi{10.1007/s00348-018-2496-1}

\bibitem[{Ern et~al(2012)Ern, Risso, Fabre, and Magnaudet}]{Ern.2012}
Ern P, Risso F, Fabre D, Magnaudet J (2012) Wake-induced oscillatory paths of
  bodies freely rising or falling in fluids. Annual Review of Fluid Mechanics
  44(1):97--121, \doi{10.1146/annurev-fluid-120710-101250}

\bibitem[{Gonz{\'a}lez et~al(1996)Gonz{\'a}lez, Resa, Ruiz, and
  Guti{\'e}rrez}]{Gonzalez.1996}
Gonz{\'a}lez C, Resa JM, Ruiz A, Guti{\'e}rrez JI (1996) Densities of mixtures
  containing n -alkanes with sunflower seed oil at different temperatures.
  Journal of Chemical {\&} Engineering Data 41(4):796--798,
  \doi{10.1021/je960053p}

\bibitem[{Hosokawa et~al(2017)Hosokawa, Nishikawa, Kogawa, and
  Takeyama}]{Hosokawa.2017}
Hosokawa M, Nishikawa Y, Kogawa M, Takeyama H (2017) Massively parallel whole
  genome amplification for single-cell sequencing using droplet microfluidics.
  Scientific reports 7(1):5199, \doi{10.1038/s41598-017-05436-4}

\bibitem[{Kang et~al(2014)Kang, Ali, .Zhang, S.Huang, Peterson, Digman,
  Gratton, and Zhao}]{Kang.2014}
Kang DK, Ali MM, Zhang K, SHuang S, Peterson E, Digman MA, Gratton E, Zhao W
  (2014) Rapid detection of single bacteria in unprocessed blood using
  integrated comprehensive droplet digital detection. Nature Communications
  5:5427, \doi{10.1038/ncomms6427}

\bibitem[{Khodaparast et~al(2013)Khodaparast, Borhani, Tagliabue, and
  Thome}]{Khodaparast.2013}
Khodaparast S, Borhani N, Tagliabue G, Thome JR (2013) A micro particle shadow
  velocimetry ($\mu$psv) technique to measure flows in microchannels.
  Experiments in Fluids 54(2):1474, \doi{10.1007/s00348-013-1474-x},
  \urlprefix\url{https://doi.org/10.1007/s00348-013-1474-x}

\bibitem[{Kinoshita et~al(2007)Kinoshita, Kaneda, Fujii, and
  Oshima}]{Kinoshita.2006}
Kinoshita H, Kaneda S, Fujii T, Oshima M (2007) Three-dimensional measurement
  and visualization of internal flow of a moving droplet using confocal
  micro-piv. Lab Chip 7:338--346, \doi{10.1039/B617391H},
  \urlprefix\url{http://dx.doi.org/10.1039/B617391H}

\bibitem[{Kobayashi et~al(2006)Kobayashi, Mori, and Kobayashi}]{Kobayashi.2006}
Kobayashi J, Mori Y, Kobayashi S (2006) Multiphase organic synthesis in
  microchannel reactors. Chemistry, an Asian journal 1(1-2):22--35,
  \doi{10.1002/asia.200600058}

\bibitem[{Kovalev et~al(2018)Kovalev, Yagodnitsyna, and Bilsky}]{Kovalev.2018}
Kovalev AV, Yagodnitsyna AA, Bilsky AV (2018) Flow hydrodynamics of immiscible
  liquids with low viscosity ratio in a rectangular microchannel with
  t-junction. Chemical Engineering Journal 352:120--132,
  \doi{10.1016/j.cej.2018.07.013}

\bibitem[{Kralj et~al(2007)Kralj, Sahoo, and Jensen}]{Kralj.2007}
Kralj JG, Sahoo HR, Jensen KF (2007) Integrated continuous microfluidic
  liquid-liquid extraction. Lab on a chip 7(2):256--263, \doi{10.1039/b610888a}

\bibitem[{Krohn et~al(2018)Krohn, Manera, and Petrov}]{Krohn.2018}
Krohn B, Manera A, Petrov V (2018) A novel method to create high density
  stratification with matching refractive index for optical flow
  investigations. Experiments in Fluids 59(4):434,
  \doi{10.1007/s00348-018-2522-3}

\bibitem[{Lang et~al(2012)Lang, Hill, Krossing, and Woias}]{Lang.2012}
Lang P, Hill M, Krossing I, Woias P (2012) Multiphase minireactor system for
  direct fluorination of ethylene carbonate. Chemical Engineering Journal
  179:330 -- 337, \doi{https://doi.org/10.1016/j.cej.2011.11.015},
  \urlprefix\url{http://www.sciencedirect.com/science/article/pii/S1385894711013945}

\bibitem[{Liu et~al(2017)Liu, Zhang, Pang, Wang, and Li}]{Liu.2017}
Liu Z, Zhang L, Pang Y, Wang X, Li M (2017) Micro-piv investigation of the
  internal flow transitions inside droplets traveling in a rectangular
  microchannel. Microfluidics and Nanofluidics 21(12):180,
  \doi{10.1007/s10404-017-2019-z},
  \urlprefix\url{https://doi.org/10.1007/s10404-017-2019-z}

\bibitem[{{Luis A. M. Rocha} and {Jo{\~a}o M. Miranda and Joao B. L. M.
  Campos}(2017)}]{LuisA.M.Rocha.2017}
{Luis A M Rocha}, {Jo{\~a}o M Miranda and Joao B L M Campos} (2017) Wide range
  simulation study of taylor bubbles in circular milli and microchannels.
  Micromachines 8(5):154, \doi{10.3390/mi8050154}

\bibitem[{Ma et~al(2014)Ma, Sherwood, Huck, and Balabani}]{Ma.2014}
Ma S, Sherwood JM, Huck WTS, Balabani S (2014) On the flow topology inside
  droplets moving in rectangular microchannels. Lab on a chip
  14(18):3611--3620, \doi{10.1039/c4lc00671b}

\bibitem[{Magnaudet and Eames(2000)}]{Magnaudet.2000}
Magnaudet J, Eames I (2000) The motion of high-reynolds-number bubbles in
  inhomogeneous flows. Annual Review of Fluid Mechanics 32(1):659--708,
  \doi{10.1146/annurev.fluid.32.1.659}

\bibitem[{Mazutis et~al(2013)Mazutis, Gilbert, Ung, Griffiths, and
  Heyman}]{Mazutis.2013}
Mazutis L, Gilbert J, Ung WL, Griffiths AD, Heyman JA (2013) Single-cell
  analysis and sorting using droplet-based microfluidics. Nature Protocols
  8:870–891, \doi{10.1038/nprot.2013.046}

\bibitem[{Miessner et~al(2008)Miessner, Lindken, and
  Westerweel}]{Miessner.2008}
Miessner U, Lindken R, Westerweel J (2008) Velocity measurements in microscopic
  two-phase flows by means of micro piv. In: Proceedings of the 6th
  International Conference on Nanochannels, Microchannels and Minichannels -
  2008, ASME, New York, NY, pp 1111--1118, \doi{10.1115/ICNMM2008-62093}

\bibitem[{Najjari et~al(2016)Najjari, Hinke, Bulusu, and
  Plesniak}]{Najjari.2016}
Najjari MR, Hinke JA, Bulusu KV, Plesniak MW (2016) On the rheology of
  refractive-index-matched, non-newtonian blood-analog fluids for piv
  experiments. Experiments in Fluids 57(6):1704,
  \doi{10.1007/s00348-016-2185-x}

\bibitem[{Park and Kihm(2006)}]{Park.2006}
Park JS, Kihm KD (2006) Three-dimensional micro-ptv using deconvolution
  microscopy. Experiments in Fluids 40(3):491--499,
  \doi{10.1007/s00348-005-0090-9}

\bibitem[{Piao et~al(2015)Piao, Han, Azad, Park, and Seo}]{Piao.2015}
Piao Y, Han DJ, Azad MR, Park M, Seo TS (2015) Enzyme incorporated microfluidic
  device for in-situ glucose detection in water-in-air microdroplets.
  Biosensors and Bioelectronics 65:220 -- 225,
  \doi{https://doi.org/10.1016/j.bios.2014.10.032}

\bibitem[{{R. G. LeBel and D. A. I. Goring}(1962)}]{LeBel.1962}
{R G LeBel and D A I Goring} (1962) Density, viscosity, refractive index, and
  hygroscopicity of mixtures of water and dimethyl sulfoxide. Journal of
  Chemical and Engineering Data

\bibitem[{Rao and Wong(2018)}]{Rao.2018}
Rao SS, Wong H (2018) The motion of long drops in rectangular microchannels at
  low capillary numbers. Journal of Fluid Mechanics 852:60--104,
  \doi{10.1017/jfm.2018.521}

\bibitem[{Saksena et~al(2015)Saksena, Christensen, and
  Pearlstein}]{Saksena.2015}
Saksena R, Christensen KT, Pearlstein AJ (2015) Surrogate immiscible liquid
  pairs with refractive indexes matchable over a wide range of density and
  viscosity ratios. Physics of Fluids 27(8):087,103, \doi{10.1063/1.4928030}

\bibitem[{Shi et~al(2019)Shi, Nie, Dong, Long, Xu, and Liu}]{Shi.2019}
Shi H, Nie K, Dong B, Long M, Xu H, Liu Z (2019) Recent progress of
  microfluidic reactors for biomedical applications. Chemical Engineering
  Journal 361:635--650, \doi{10.1016/j.cej.2018.12.104}

\bibitem[{Sinzato et~al(2017)Sinzato, {Sousa Dias}, and Cunha}]{Sinzato.2017}
Sinzato YZ, {Sousa Dias} NJ, Cunha FR (2017) An experimental investigation of
  the interfacial tension between liquid-liquid mixtures in the presence of
  surfactants. Experimental Thermal and Fluid Science 85:370--378,
  \doi{10.1016/j.expthermflusci.2017.03.011}

\bibitem[{Song et~al(2006)Song, Chen, and Ismagilov}]{Song.2006}
Song H, Chen DL, Ismagilov RF (2006) Reactions in droplets in microfluidic
  channels. Angewandte Chemie International Edition 45(44):7336--7356,
  \doi{10.1002/anie.200601554},
  \urlprefix\url{https://onlinelibrary.wiley.com/doi/abs/10.1002/anie.200601554},
  \eprint{https://onlinelibrary.wiley.com/doi/pdf/10.1002/anie.200601554}

\bibitem[{Tanimu et~al(2017)Tanimu, Jaenicke, and Alhooshani}]{Tanimu.2017}
Tanimu A, Jaenicke S, Alhooshani K (2017) Heterogeneous catalysis in continuous
  flow microreactors: A review of methods and applications. Chemical
  Engineering Journal 327:792 -- 821, \doi{10.1016/j.cej.2017.06.161}

\bibitem[{Weast(1989)}]{Weast.1989}
Weast RC (ed)  (1989) CRC handbook of chemistry and physics: A ready-reference
  book of chemical and physical data, 70th edn. {CRC Press}, Boca Raton

\bibitem[{Wilkinson(1972)}]{Wilkinson.1972}
Wilkinson M (1972) Extended use of, and comments on, the drop-weight
  (drop-volume) technique for the determination of surface and interfacial
  tensions. Journal of Colloid and Interface Science 40(1):14--26,
  \doi{10.1016/0021-9797(72)90169-5}

\bibitem[{Wolf et~al(2015)Wolf, Hartmann, Bertolini, Schemberg, Grodrian,
  Lemke, Förster, Kessler, Hänschke, Mertens, Paus, and Lerchner}]{Wolf.2015}
Wolf A, Hartmann T, Bertolini M, Schemberg J, Grodrian A, Lemke K, Förster T,
  Kessler E, Hänschke F, Mertens F, Paus R, Lerchner J (2015) Toward
  high-throughput chip calorimetry by use of segmented-flow technology.
  Thermochimica Acta 603:172 -- 183, \doi{10.1016/j.tca.2014.10.021}, chip
  Calorimetry

\bibitem[{Wright et~al(2017)Wright, Zadrazil, and Markides}]{Wright.2017}
Wright SF, Zadrazil I, Markides CN (2017) A review of solid--fluid selection
  options for optical-based measurements in single-phase liquid, two-phase
  liquid--liquid and multiphase solid--liquid flows. Experiments in Fluids
  58(9):357, \doi{10.1007/s00348-017-2386-y}

\bibitem[{Zhao and Middelberg(2011)}]{Zhao.2011}
Zhao CX, Middelberg AP (2011) Two-phase microfluidic flows. Chemical
  Engineering Science 66(7):1394--1411, \doi{10.1016/j.ces.2010.08.038}

\end{thebibliography}

\end{document}